\documentclass[prd,preprintnumbers,showpacs,nofootinbib, 11pt]{revtex4}%{revtex4} ,twocolumn
\usepackage[english]{babel}
\usepackage{graphicx}% Include figure files
\usepackage{subfig}
\usepackage{hyperref}
\usepackage{dcolumn}% Align table columns on decimal point
\usepackage{amsmath,amssymb,indentfirst,color}
\usepackage{slashed}
\usepackage{hyperref}
\usepackage{amsfonts}
\usepackage{feynmp}

\def\bea{\begin{eqnarray}}
\def\eea{\end{eqnarray}}
\def\beq{\begin{equation}}
\def\eeq{\end{equation}}
\def\to{\rightarrow}
\newcommand{\beas}{\begin{eqnarray*}}
\newcommand{\eeas}{\end{eqnarray*}}
\newcommand{\ba}{\begin{array}}
\newcommand{\ea}{\end{array}}

\newcommand{\gae}{\begin{array}{c}\,\vspace{-0.5em}\hspace{-0.2em}\sim\vspace{-1.7em}\\>
\end{array}}
\newcommand{\lae}{\begin{array}{c}\,\vspace{-0.5em}\hspace{-0.2em}\sim\vspace{-1.7em}\\<
\end{array}}

\begin{document}

\rightline{\vbox{\small\hbox{\tt NITS-PHY-2013002} }}

\title{Proton Compton scattering in a unified proton-$\Delta^+$ Model}

\author{ Yun Zhang and Konstantin Savvidy}
\affiliation{Department of Physics, Nanjing University, Nanjing, China}
\begin{abstract} 
We develop a field-theoretic model for the description of proton Compton scattering in which the proton and its excited state, the $\Delta^+$ resonance, are described as part of one multiplet with a single Rarita-Schwinger wavefunction. In order to describe the phenomena observed, it is necessary to incorporate both minimal and non-minimal couplings. The minimal coupling reflects the fact that the $\Delta^+$ is a charged particle, and in this model the minimal coupling contributes also to the $\gamma N \Delta$ magnetic transition. The non-minimal couplings consist of  five electromagnetic form-factors, which are accessed at fixed and vanishing  momentum-transfer squared with real photons in Compton scattering experiments, therefore it is possible to extract a rather well-determined set of optimal parameters which reasonably well fit the data in the resonance region 140-450 MeV. The crucial parameter which determines the $\gamma N\Delta$ transition amplitude and therefore the height of the resonance peak is equal to $3.66 \pm 0.03$, in units of $\mu_N$. We find that this parameter also primarily determines the contributions to magnetic polarizability in this model.
In the low-energy region up to 140 MeV, we separately fit the electric and magnetic polarizabilities, while keeping the other parameters fixed and obtain values in line with previous approaches. The basic model is then extended with insights gained in the traditional approaches, namely incorporating the sigma-meson channel with the currently favored parameters, and the pion vertex corrections.

\end{abstract} 
\date{\today}
\pacs{12.39.Fe, 13.60.Fz, 14.20.Dh}

\maketitle

\section{Background and Introduction}

%During these two decades much progress has been made in experimental techniques in measuring proton Compton scattering\cite{Hallin1993,Federspiel1991,MacGibbon1995,Leon2001}, Theories having been developed in this direction includes dispersion calculations\cite{Petrunkin1981,Lvov1993,Holstein1994,Baldin1960,Lvov1997}, the chiral bag model\cite{Weiner1985}, constituent quark model\cite{Capstick1992}, effective-Lagrangian models\cite{Pascalutsa1995,Scholten1996}, the chiral perturbation theory\cite{Bernard1992,Bernard1995}, and soliton models\cite{Scoccola1990}. Among them, Baldin dispersion relation\cite{Baldin1960} is one of the most successful. 
%Previous works have taken the Delta baryon into consideration either as an implicit\cite{Bernard1992,Bernard1995} or explicit\cite{Jenkins1992,Hemmert1997} degree of freedom\footnote{maybe the cites should be removed here}. Previous works use the result derived in\cite{Powell1949}, 

%describe the current experimental and theoretical status of polarizabilities in more detail and add references
%obscure
Proton is the particle which makes up the greatest fraction of the matter in the visible universe and its properties have been extensively studied. Nevertheless it still holds some mysteries, among them the physical origin 
of the electric and magnetic polarizabilities, 
right behind the more fundamental electromagnetic properties of the proton as are the electric charge and magnetic moment. Experiments have been done since the 60's to characterize and measure  the electromagnetic properties of the proton using fixed-target Compton scattering \cite{Baranov1974, Zieger92, MacGibbon1995, Hunger97, Federspiel1991, Wada:1981ab} . In the recent two decades, high quality proton Compton scattering data in the first $\Delta$(1232MeV) resonance region have been obtained at Saskatchewan \cite{Hallin1993},  by LEGS Collaboration \cite{Blanpied01} and at Mainz MAMI \cite{Olmos2001,MAMI2001}. Also, in the higher energy region where some recent good data is available due to the Hall A Collaboration at Jefferson Lab \cite{Danagoulian2007}.
These most recent and precise  experiments have determined the static values of the  electric and magnetic polarizabilities, see for example \cite{Wright2004, Schumacher2005}, particularly the sign and value of the magnetic polarizability which for a long time remained shrouded in uncertainty, but see \cite{Krupina:2013dya} for a new proposal for additional experiments on this. Also, high precision value was obtained  for the spin-polarizability \cite{Olmos2001, MAMI2001} which appears in the expansion of the scattering amplitudes to the third order in momentum.

%available In recent yearsstill has some important parameters not precisely measured and not well predicted from theory. 
%continue to attract much investigation in recent years. The electric and magnetic polarizabilities measure the induced electric and magnetic dipoles of the proton in static electric and magnetic fields. 
%Experimentally, it is convenient to study the electromagnetic properties of the proton using real Compton scattering. Real photons, like the virtual ones present in the static electric and magnetic fields, also polarize the proton and affect the scattering amplitudes in the second order in the frequency. 
%

Experimentally, it is certainly also possible to measure polarization asymmetries as a function of angle and energy, for example in the last experiment done at the venerable Yerevan accelerator \cite{al:1993aa} and in the first resonance region by the LEGS collaboration \cite{Blanpied:1996yh}. These data can well be used to discriminate any theoretical model as a strong cross-check, once the basic parameters of the model have been well-determined.

%The most phenomenologically successful description until now has been that based on dispersion theory \cite{Baldin1960, GellMann:1954db, Guiasu:1978ak, Lvov:1979zd, Lvov1997, Hearn:1962zz, Pfeil:1974ib, Lvov:1980wp, L'vov:1996xd, Drechsel:1999rf, Pasquini2007}, in which scattering amplitudes are constructed as analytic functions of momenta and coefficients are chosen based on elaborate understanding of these low energy scattering processes.
%Another alternative is to use an effective theory \cite{Scholten1996, Kondratyuk2001, Feuster:1998cj, Pascalutsa1995}, such as the chiral perturbation theory \cite{Pagels1974, Weinberg1978, Gasser1983, Gasser1987}.

 The process in the photon low energy range up to 140 MeV is dominated by contributions due to the anomalous magnetic moment as well as the polarizabilities of the nucleon. Of these additional contributions, anomalous magnetic moments contribute to the amplitude already at the linear order while polarizability starts out at the second order, thus cross-section can grow at first quadratically and then quartically with energy. This is in contrast to the minimal coupling in QED, where the Klein-Nishina cross-section is essentially constant in the relevant energy range.

Fundamental results on scattering of light on particles of spin 1/2 with anomalous magnetic moment were obtained by Powell, Low, Gell-Mann and Goldberger, \cite{Powell1949, GellMann1954,  Low1954}.  Early theoretical progress was driven by the phenomenologically very successful dispersion theory approach \cite{Baldin1960, Hearn:1962zz, Pfeil:1974ib, Guiasu:1978ak, Petrunkin1981, Lvov1993, Lvov93, Lvov1997, Lvov:1979zd,  Lvov:1980wp, L'vov:1996xd, Drechsel:1999rf, Pasquini2007}, see also the latest excellent review in \cite{Schumacher2013}. This approach was supplemented by insights gained from considering pion-vertex corrections and a multitude of other improvements such as those in \cite{Feuster:1998cj, Jenkins1992, Kondratyuk2001, Schumacher2005,  Dattoli1977}. 

On the other hand, within the purely field-theoretic  chiral-Lagrangian paradigm \cite{Weinberg1978, Gasser1983}, the development of the promising approach of Peccei \cite{Peccei1968, Peccei1969} was held up by difficulties in the field theory of the spin 3/2 $\Delta^+$ particle. The most egregious of these pathologies {had} been resolved in \cite{Benmerrouche:1989uc}. This better understanding of the theoretical requirements on the $\Delta$ propagator led to the paper of  Pascalutsa and Scholten \cite{Pascalutsa1995} in which the first workable field-theoretical model for proton Compton scattering which incorporated the contribution of the resonance was constructed. There, it was argued that the virtual spin 1/2 degrees of freedom present in the standard propagator do play a role in the Compton scattering amplitude. As we shall see, the present work's approach most directly descends from this model of Pascalutsa and Scholten and the subsequent recent developments in \cite{Scholten1996, 09053861, Pascalutsa:1999zz, Lensky2010, McGovern:2012ew, Lensky:2012ag,Griesshammer:2012we, Phillips:2012xa}.

 %Pascalutsa and Scholten also discussed in detail the spin 3/2 propagator which is necessary for calculating the contributions of the $\Delta^+$ exchange diagrams, finding that better agreement is found if spin 1/2 contributions are kept in the propagator, which led them to conclude that the spin 1/2 contributions may be remnants of some high-mass excited states of the nucleon.
 
A separate development was the proposal to rid the theory of spin 3/2 particles of pathologies that stem from superluminal solutions by Ranada and Sierra in \cite{Ranada:1980yx}. There, it was found that taking a Rarita-Schwinger multiplet of a physical spin 3/2 and a  physical spin 1/2 particle (of different mass), would result in an acceptable wave equation even when minimally coupled to the electromagnetic field. A  detailed investigation of the structure of poles in the propagator revealed that additional restrictions on the Ranada and Sierra equations  result in a unitary theory with positive definite residues at the Feynman poles, taking into account the location of the pole above or below the real axis \cite{Kostas10}.  

In the present paper we make use of the propagator of  \cite{Kostas10}, in combination with the observation of Pascalutsa and Scholten that the spin 1/2 degrees of freedom may be due to another baryon, and propose a model where this spin 1/2 mode is interpreted as the proton, such that the proton and the $\Delta$ are together described by a single multi-component wave function of Rarita-Schwinger. $\Delta^+$ has the same quark constitution (uud) as the proton and is only slightly, less than 300MeV heavier than the proton.  The only difference between the $\Delta^+$ and proton is the alignment of the spins of these quarks. In the proton, the d-quark spin is anti-aligned and makes the total proton spin $\frac{1}{2}$, while in $\Delta^+$ all three quarks are aligned, making $\Delta^+$ spin $\frac{3}{2}$. This makes it very natural to consider the possibility of a unified description, see Section \ref{sec:lagr} for the Lagrangian and more details.

\begin{figure}[htdp]
%\setlength{\unitlength}{1cm}
%\center{
\includegraphics[height=5cm,width=10cm]{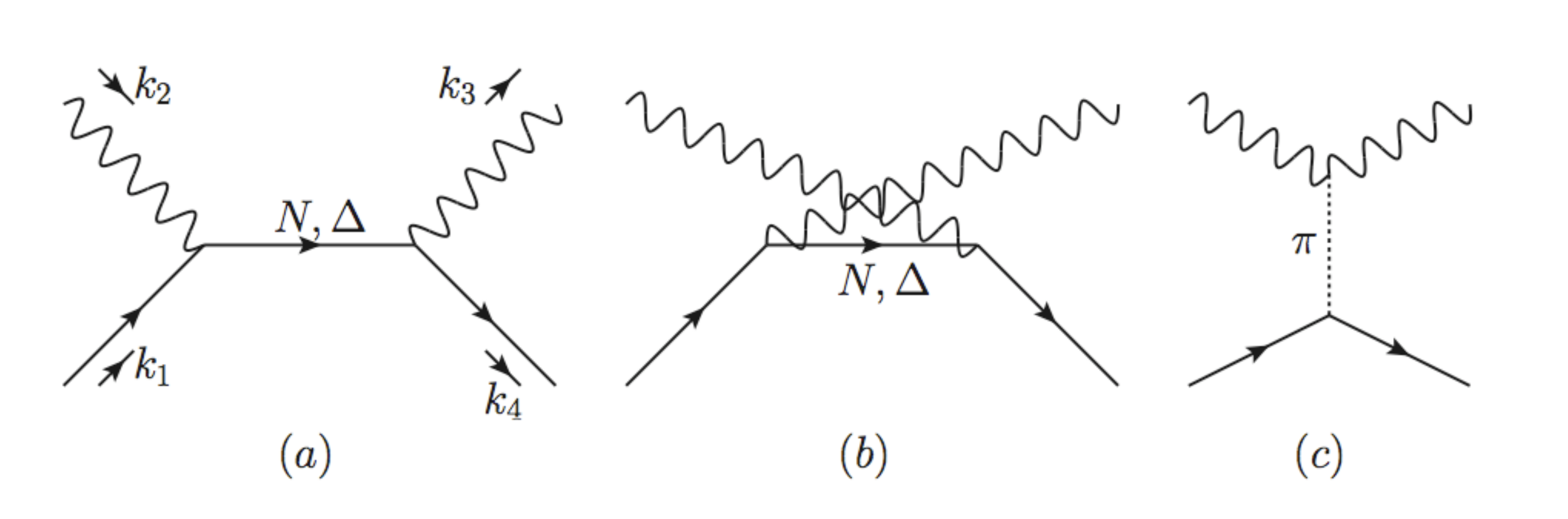}\raisebox{1mm}{\includegraphics[height=4.4cm,width=2.6cm]{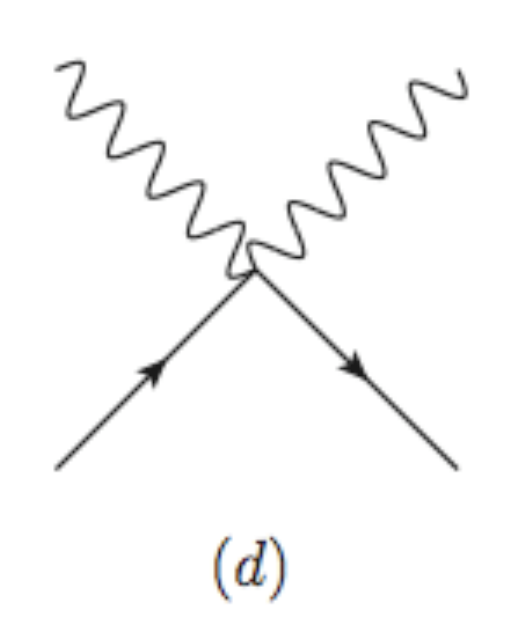}}
\caption{Tree-level Feynman diagrams for proton Compton scattering. In (a) and (b) the intermediate particle is proton or $\Delta^+$, and in (c) the $\pi^0$ meson is exchanged. (d) is the diagram for the contact interaction in (\ref{poleffLag}). For appropriate photon incident energy, the intermediate $\Delta^+$ is approximately on-shell, and around this energy, there is the characteristic peak in the cross-section which is dominated by the $\Delta^+$ contribution.
}
\label{fig2}
\end{figure}

This hypothesis confers several benefits. Low energy proton Compton scattering occurs mainly  through the proton and $\Delta^+$ {in the s- and u- channel} and exchange of pions in the t-channel,  see Fig. \ref{fig2}. %Pion exchange amplitude is easy to calculate and involves little dispute. 
$\Delta^+(1232{\mbox MeV}, J^P=\frac{3}{2}^+)$, the lightest baryon resonance, appears in intermediate state.
 Furthermore, $\Delta^+$ is only slightly heavier than the proton, so that this contribution is not suppressed even at the lowest energies compared with the exchange of the proton alone.
In our model, because of the unified description, the s/u-channel proton and $\Delta^+$ contributions can be calculated simultaneously as in Fig. \ref{fig2} a) and b) instead of adding up the four separate contributions. 

The second benefit is that in this model it is possible to avoid the introduction of a large number of arbitrary parameters. In Section \ref{sec:nonmin} (see also \cite{chensav}), we present in detail the five electromagnetic form-factors which are possible for a spin 3/2 particle, these couple the momentum-independent fermion bilinears directly to the EM field strength. Three of these have  clear physical interpretation as the magnetic moments of the proton and the $\Delta^+$, and the strength of the M1 magnetic transition between $N$ and $\Delta^+$. One of the remaining two parameters contributes a purely imaginary part to some amplitudes and this is disfavored by data, so that it can be safely set to zero (see Section \ref{sec:fit}).

 In Section \ref{sec:EMquantity}, we discuss the N-N, $\Delta^+$-$\Delta^+$, N-$\Delta^+$ transition matrices and derive the formulae of the proton and $\Delta^+$ magnetic moments.
    In Section \ref{sec:crosssec}, we calculate proton Compton scattering cross section and express electric and magnetic polarizabilites in terms of the coefficients of the non minimal interactions, and also analyze the behavior of the amplitudes around the $\Delta^+$ pole. %, using the known expression of extracting polarizabilities from cross section, see eq.(\ref{polarizability}). 
{We fit the Compton scattering data in Section \ref{sec:fit}}, and extract polarizabilities using expressions derived in Section \ref{sec:crosssec}. Since one linear combination of the coefficients of the non minimal interactions gives the $\Delta^+$ magnetic moment, our best-fit parameter set also provides a prediction of the $\Delta^+$ magnetic moment, but we conservatively interpret it as an upper bound. 
The interactions in Section \ref{sec:nonmin} can also be used to calculate the $\Delta^+\to N+\gamma$ decay process (Appendix \ref{sec:decayamp}), this acts as a useful cross-check on our model. 
%{Accommodating $\Delta^+\to p+\gamma$ decay width and $\Delta^+$ magnetic moment, at the same time of extracting polarizabilites from Compton scattering cross section, ... }
% is an advantage of our work. 
%Finally in Section 7 we conclude.
%the sign of $\pion^0-\gamma-\gamma$ vertex as a by-product?

%Since the process involves spin 3/2 $\Delta^+$, a spin 3/2 field theory is needed. %One widely used spin 3/2 theory is Rarita-Schwinger theory, although known to suffer from superluminal propagation problem when coupled minimally to the electromagnetic field. 
%Peccei \cite{Peccei1968, Peccei1969} introduced a workable effective theory of the $\Delta$ including the lowest order effective couplings to the photon. Pascalutsa and Scholten \cite{Pascalutsa1995} extended the Peccei approach with additional allowed forms of the $\gamma N \Delta^+$  interactions and fit then current experimental data to their model. %., used Rarita-Schwinger theory to describe $\Delta^+$ and proposed $p-\Delta^+-\gamma$ interaction vertex. 
%In models of this kind the proton Compton scattering is represented as the sum of s,u-channel proton exchange, s,u-channel $\Delta^+$ exchange and t-channel $\pi^0$ exchange contributions, see Fig. 1 of \cite{Pascalutsa1995}.

{\section{Lagrangian and the electromagnetic interactions}\label{sec:min}}
\label{sec:lagr}
%Recently in \cite{Kostas10}, a model is proposed in which one can get rid of the superluminal problem of the original Rarita-Schwinger theory at the cost of allowing an additional spin 1/2 physical particle, see also \cite{Ranada:1980yx}. 
A spin 3/2 field is represented as a field with both a Lorentz index and a Dirac index. The Lagrangian in \cite{Ranada:1980yx,Kostas10} is:
\bea
&&{\mathcal L}= - {\bar \psi}_\lambda \, [p_\mu \, \Gamma^{\mu\lambda}{}_\rho-m \, \Theta^\lambda{}_\rho] \, \psi^\rho,\nonumber\\
&&\Gamma^{\mu\lambda}{}_\rho=\gamma^\mu \, \eta^\lambda{}_\rho+\xi \, (\gamma^\lambda \,  \eta^\mu{}_\rho+\gamma_\rho \,  \eta^{\lambda\mu})+\zeta \, \gamma^\lambda \, \gamma^\mu \, \gamma_\rho,\nonumber\\
&&\Theta^\lambda{}_\rho=\eta^\lambda{}_\rho- z  \, \gamma^\lambda \, \gamma_\rho,\nonumber\\
&&\xi=2\, z-1,\ \ \ \ \zeta= 6\, z^2-4\,z+1 , % \frac{3\, \xi^2+2\, \xi+1}{2},
\eea
in which the Dirac indices are suppressed. The solutions of the corresponding wave equation are all transverse, $p_\mu \,\psi^\mu =0$, and consist of a spin 3/2 with mass m and a spin 1/2 component with mass $M=\frac{m}{6z-2}$, compared to a pure spin 3/2 field in the original Rarita-Schwinger theory. We identify the spin 3/2 and 1/2 component as $\Delta^+$ and proton respectively and thus unify them in one theory. This unification is very natural since proton and $\Delta^+$ have the same quark constituents and can transform from one into the other by absorbing or emitting a photon or even a neutral pion (which carries no charge or spin). The mass splitting between the $N$ and $\Delta$ is less than 300 MeV. The model allows to adjust the ratio of the proton to the Delta mass by choice of the $z$ parameter.

The propagator with the specific arragement of the poles is \cite{Kostas10}:
% \footnote{{ZY: why -i?   K: do you disagree?}} 
\bea
-i S(p)&=
% &\frac{2m \Pi^+(p,m)\Pi_3(p)}{p^2-m^2+i\epsilon}-\frac{M}{(3z-1)^2}\frac{\Pi^+(p,M)\Pi_{11}(p)}{p^2-M^2- i\epsilon}\\
% &+&\frac{1}{m-4zm}(\frac{3z^2}{3z-1}\Pi_{11}(p)-z\Pi_{12}(p)-z\Pi_{21}(p)+(3z-1)\Pi_{22}(p)),\nonumber\\
 & \frac{(\slashed{p}+m)\, \Pi_3  }{p^2-m^2+i\,\epsilon}  
- \frac{(\slashed{p}+M) \, \Pi_{11} }{p^2-M^2 - i\,\epsilon} \, \frac{2M^2}{m^2} \nonumber\\ % \frac {1} {2\,(3\,z-1)^2}
&+&  %\frac{(1-3z)}{(1-4z)\,m} \, 
\left[ \vspace{30pt} \Pi_{22} - (\Pi_{21}+\Pi_{12})\,/B +  \Pi_{11}\,3/B^2 \right] \, \frac{3}{2 \,(M +2 \, m)}~~,\nonumber\\
B &=& \frac{3\,m}{2 \, M +m} ~~.%(3z-1)/z
\eea
where the standard spin projection operators $\Pi$ can be found for example in \cite{Benmerrouche:1989uc}.

The minimal electromagnetic interaction is usually derived by substituting $p_\mu \to p_\mu + e \, A_\mu$. The interaction Lagrangian is:
\beq
{\mathcal L}_I=e \, {\bar \psi}_\lambda  \, \Gamma^{\mu\lambda}{}_\rho  \, \psi^\rho \, A_\mu~.
\eeq
Ward identity is satisfied:
\beq
-i  \, k_\mu  \, \Gamma^{\mu\lambda}{}_\rho=S^\lambda{}_\rho(p+k)^{-1}-S^\lambda{}_\rho(p)^{-1}~,
\label{wardid}
\eeq
%({introduce the propagator here?})
and as we shall see in the next Section there is only a finite number of additional possibilities of electromagnetic coupling which satisfy gauge invariance and the Ward identity.

\subsection{Non-Minimal Electromagnetic Interactions}
\label{sec:nonmin}
For phenomenological application to proton Compton scattering, minimal interaction alone does not suffice. A well known fact for Dirac theory is that it allows for two electromagnetic form factors one of which is charge and the other describes the anomalous magnetic moment. %minimal electromagnetic interaction $e{\bar u}\gamma^\mu u A_\mu$ cannot describe anomalous magnetic moment, and one should add non-minimal interaction $\frac{i e}{2m} F_2 {\bar u}\sigma^{\mu\nu} u q_\nu A_\mu$ to describe anomalous magnetic moment, the coefficients known as form factors. 
%Spin 3/2 fields have more complicated electromagnetic properties, so 
We add these as yet undetermined non-minimal interactions to the vertex:
\beq
{\tilde \Gamma}^{\mu\lambda}{}_\rho= \Gamma^{\mu\lambda}{}_\rho+ \frac{i}{2M} \, \sum\limits_{n} F_n(k^2) \, (\Gamma_n{})^{\mu\lambda}{}_\rho,
\label{genvertex}
\eeq
where the $F_n(k^2)$ are form factors. If amplitudes are to be gauge invariant, the
Ward identity (\ref{wardid}) should still hold. For that, it is sufficient to set $k_\mu (\Gamma_n){}^{\mu\lambda}{}_\rho=0$ and thus $\Gamma_n$ should be of the form:
\beq
(\Gamma_n)^{\mu\lambda}{}_\rho=(\Sigma_n)^{\mu\nu\lambda}{}_\rho \, k_\nu,
\label{nonminvertex}
\eeq
where $(\Sigma_n)^{\mu\nu\lambda}{}_\rho$ is antisymmetric in $\mu$ and $\nu$.

Antisymmetric tensors live in the $(1,0)\oplus(0,1)$ representation {of Lorentz group}, and we can count the number of these representations in the product representation of the two matter fields. Representation for $\psi_\lambda$(or ${\bar \psi}_\lambda$) is a product of that for a vector field and that for a spinor field:
\beq
(\frac{1}{2},\frac{1}{2})\otimes[(\frac{1}{2},0)\oplus(0,\frac{1}{2})]=(1,\frac{1}{2})\oplus(0,\frac{1}{2})\oplus(\frac{1}{2},0)\oplus(\frac{1}{2},1).
\eeq
The vertexes live in the tensor product of the above reducible representations, and
\bea
&&[(1,\frac{1}{2})\oplus(0,\frac{1}{2})\oplus(\frac{1}{2},0)\oplus(\frac{1}{2},1)]\otimes[(1,\frac{1}{2})\oplus(0,\frac{1}{2})\oplus(\frac{1}{2},0)\oplus(\frac{1}{2},1)]\nonumber\\
&& \supset 5\ [(1,0)\oplus(0,1)]
\eea
This tells us there are five antisymmetric tensors and we have been able to explicitly construct them as:
\bea
&&(\Sigma_1)^{\mu\nu\lambda}{}_\rho=-\frac{1}{2}\tau^{\mu\nu\lambda}{}_\rho\ ,\nonumber\\
&&(\Sigma_2)^{\mu\nu\lambda}{}_\rho=\sigma^{\mu\nu}\eta^\lambda{}_\rho\ ,\nonumber\\
&&(\Sigma_3)^{\mu\nu\lambda}{}_\rho=-\frac{1}{9}\gamma^\lambda\sigma^{\mu\nu}\gamma_\rho\ ,\nonumber\\
&&(\Sigma_4)^{\mu\nu\lambda}{}_\rho=\frac{1}{12}(\gamma^\lambda\gamma^\mu \eta^\nu{}_\rho-\gamma^\lambda\gamma^\nu \eta^\mu{}_\rho+\gamma^\mu\gamma_\rho \eta^{\nu\lambda}-\gamma^\nu\gamma_\rho \eta^{\mu\lambda})\ ,\nonumber\\
&&(\Sigma_5)^{\mu\nu\lambda}{}_\rho=\frac{-i}{12}(\gamma^\lambda\gamma^\mu \eta^\nu{}_\rho-\gamma^\lambda\gamma^\nu \eta^\mu{}_\rho-\gamma^\mu\gamma_\rho \eta^{\nu\lambda}+\gamma^\nu\gamma_\rho \eta^{\mu\lambda}),
\label{antisymtensor}
\eea
%a11=1/4F1,c11=-1/9F3,d2=F5/12,d1=F4/12, pay attention to the change of the sign in the definition of $\tau$ to respect commutation relations as spin 1 angular momentum operator, thus replace my all to 1/4F1 instead of minus it. my definition of \sigma^{\mu\nu} is not the accepted one with an additional factor of 1/2: it should be sigma=i/2(\gamma\gamma-\gamma\gamma). So together I should change and give the definition of tau and sigma and add a minus sign to \Sigma_1, add a factor of 1/2 to \Sigma_2, \Sigma_3 and \Sigma_6, as done above without rescale F1,F2,F3, F6.
%all F1 to F6 can be rescaled by 1/2 and the coefficients in \Sigma_1 to \Sigma_6 rescaled to -i/2/M,i/M,-i/9/M,i/12M,1/12M,i/M^3
where $\tau$ and $\sigma$ are generators of the Lorentz transformation for spin 1 and 1/2 respectively, $\tau^{\mu\nu\lambda}{}_\rho=i(\eta^{\mu\lambda}\eta^\nu{}_\rho-\eta^{\nu\lambda}\eta^\mu{}_\rho)$ and $\sigma^{\mu\nu}=\frac{i}{4}(\gamma^\mu\gamma^\nu-\gamma^\nu\gamma^\mu)$. % The $\gamma$ matrixes are in Weyl representation. 
The coefficients are normalized such that the form factors $F_i$ enter with equal {weight} in proton magnetic moment in eq.(\ref{magmoment}).

These tensors satisfy the requirement of Hermiticity:%\footnote{{when m is complexified, the hermiticity is violated.}}
\beq
[{\bar \psi}(p_1)_\lambda ~ \Sigma_i^{\mu\nu\lambda}{}_\rho  ~ \psi^\rho(p_2) ~(p_1-p_2)_\nu]^\dagger = - {\bar \psi}(p_2)_\lambda  ~ \Sigma_i^{\mu\nu\lambda}{}_\rho  ~  \psi^\rho(p_1) ~ (p_2-p_1)_\nu,
\eeq
which implies that the form factors  $F_i(k^2)$ are real.

%However, the above tensors are not enough for description of spin 3/2 form factors. 
At higher order in momenta there are a small number of {additional possibilities}. The pure spin 3/2 field has three form factors other than charge \cite{chensav}, while only the first and second ones in eq.(\ref{antisymtensor}) contribute for pure spin 3/2, because $\gamma_\rho \, \psi^\rho=0$ for spin 3/2 solutions. %\footnote{The form of the interaction terms are different here and in the reference, but for on-shell spin 3/2 states, we can derive a relation between the form factors here and in the reference, and express the several electromagnetic quantities using the form factors here.} 
 Thus we add one more tensor:
%\footnote{give the relation between these form factors and the electric dipole, magnetic quadrupoles etc.?}
\beq
(\Sigma_6)^{\mu\nu\lambda}{}_\rho=\frac{1}{M^2} \,  k^\lambda \, \sigma^{\mu\nu} \, k_\rho
\label{sixthtensor}
\eeq

The form factors $F_i(k^2)$ which appear as coefficients in eq.(\ref{genvertex}) are scalar functions of momentum transferred squared $k^2=(p_1-p_2)^2$. For real Compton scattering and Delta decay $\Delta^+\to p+\gamma$, the photon is on-shell $k^2$=0, so these form factors are taken to be constants in what follows. %\footnote{The coefficient for the minimal interaction is also a function of $k^2$, which equal 1 when $k^2=0$, because electric charge is 1(e).}

% At higher order in momenta there is a small number of additional possible form factors \cite{chensav}, which we do not incorporate in this work because we are interested primarily in low-energy phenomenology in the $\Delta$ resonance region.

\subsection{Bare Polarizability Effective Lagrangian}

Expanding Compton scattering cross section at low energies, static polarizabilities $\bar\alpha$ and $\bar\beta$ first enter at second order:
\beq
\frac{d \sigma}{d \Omega_{\mbox{\tiny lab}}}=\left(\frac{d \sigma}{d \Omega}\right)_{\mbox{\tiny Powell}}-\frac{e^2 \omega^2}{4 \pi M}(\frac{\bar\alpha+\bar\beta}{2}(1+\cos\theta)^2+\frac{\bar\alpha-\bar\beta}{2}(1-\cos\theta)^2)+\mathcal O(\omega^3).
\label{polarizability}
\eeq
Thus, polarizabilities are here defined in the way standard in the literature, by comparing the theoretical predictions and experimental data to the Powell cross section $\left(\frac{d \sigma}{d \Omega}\right)_{\mbox{\tiny Powell}}$, which is the differential cross section of a Dirac point particle with anomalous magnetic moment included \cite{Powell1949, Low1954, GellMann1954}. A different definition would result if the  Klein-Nishina result for the Dirac point particle without anomalous magnetic moment was taken as the basis for comparison. Such difference has sometimes led to confusion in the literature, but has been satisfactorily resolved by separating the contributions due to the anomalous magnetic moment. %The expansion of the Powell cross section to second order can be found in \cite{Bornpart}. 
Likewise, the non-minimal interaction vertices presented in this section contribute to the effective polarizabilities $\bar\alpha$ and $\bar\beta$ as we shall see in Section \ref{sec:crosssec}. % through propagation of an internal particle, which formular will be given in the next section. 
In addition to these vertices, we may include also the effective 4-point contact interactions that can contribute directly to the polarizabilities. Inspired by the effective Lagrangian proposed in \cite{0611327}, we include the following interaction Lagrangian to model ``bare" polarizability:
\beq
{\cal L}_{\mbox{\tiny pol}}=\frac{i\pi}{M}~(\bar\psi_\lambda \, \Gamma^{\mu\lambda}{}_\rho  \,\partial_\nu  \,\psi^\rho-\partial_\nu  \,\bar\psi_\lambda  \,\Gamma^{\mu\lambda}{}_\rho \psi^\rho)(\alpha_B ~ F_{\mu\rho} \, F^{\rho\nu}+\beta_B ~  \tilde{F}_{\mu\rho} \, \tilde{F}^{\rho\nu}).
\label{poleffLag}
\eeq
This Lagrangian is not unique, but other candidates contribute identically to the cross section up to the second order in the energy  of the incident photon. %, since they should obey eq.(\ref{polarizability}).

The two coefficients $\alpha_B$ and $\beta_B$ we call bare polarizabilities. The contribution of this effective Lagrangian to the lab frame Compton scattering amplitudes at second order of photon energy is:
\beq
{\cal A}_{\rm pol}=4\pi \, \alpha_B  \, \omega \omega'  \, {\vec \epsilon}'\cdot {\vec\epsilon}+4\pi \, \beta_B  \, {\vec\epsilon}'\times{\vec k}'\cdot {\vec\epsilon}\times{\vec k}+\mathcal O(\omega^3).
\eeq
The contribution of $\alpha_B$ and $\beta_B$ to the cross section is of the form in eq.(\ref{polarizability}). In the 
low energy limit where the proton is at rest before and after, and the photon frequency tends to zero, the corresponding Hamiltonian is: 
\beq
{\cal H}_{\mbox{\tiny pol}}=-2\pi(\alpha_B |\vec{E}|^2+\beta_B |\vec{H}|^2),
\label{barepol}
\eeq
in agreement with expectations, see e.g. \cite{Lvov93}.

At higher orders in momenta it is possible to define and extract more general polarizabilities  \cite{Ashley04}, such as the spin polarizabilities at cubic order.

\section{Magnetic Moments and the $\gamma\,N\,\Delta^+$ Transition Matrix}\label{sec:EMquantity}
To calculate the magnetic moments, let  $A^\mu=(0,A_x,0,0)$ and $\vec{p}_1-\vec{p}_2=q\hat{z}$, with $\vec{p}_1\to 0,\vec{p}_2\to 0$.
\bea
&&e \, \bar{u}_2(p_1,\sigma_1)  \, \tilde{\Gamma}^\mu   \, u_2(p_2,\sigma_2)   \, A_\mu = 
                  2  \,M   \, \mu_p  \left(J_y^{\left(\frac{1}{2}\right)}\right)_{\sigma_1\sigma_2}(-i q A_x)+\mathcal O(q^2),\nonumber\\
&&e \, \bar{u}_4(p_1,\sigma_1)  \, \tilde{\Gamma}^\mu   \, u_4(p_2,\sigma_2)   \, A_\mu=
                          -2  \,m   \, \mu_{\Delta^+}  \left(J_y^{\left(\frac{3}{2}\right)}\right)_{\sigma_1\sigma_2}(-i \,q \, A_x)+\mathcal O(q^2).
\label{magmom}
\eea
In this equation, $u_2$ and $u_4$ are the spin 1/2 and spin 3/2 solutions of the wave equation and $\vec{J}^{\left(\frac{1}{2}\right)}$ and $\vec{J}^{\left(\frac{3}{2}\right)}$ are standard quantum mechanical spin operators for spin 1/2 and 3/2. 
%$\vec{J}^{\left(\frac{1}{2}\right)}=\frac{\vec{\sigma}}{2}$, where $\vec{\sigma}$ are Pauli matrixes. For spin 3/2,
%\begin{eqnarray*}
%&&J_x=\frac{1}{2}\left (\begin{array}{cccc} 0 & \sqrt{3} & 0 & 0 \\ \sqrt{3} & 0 & 2 & 0 \\ 0 & 2 & 0 & \sqrt{3} \\ 0 & 0 & \sqrt{3} & 0 \end{array} \right ),J_y=\frac{i}{2}\left (\begin{array}{cccc} 0 & -\sqrt{3} & 0 & 0 \\ \sqrt{3} & 0 & -2 & 0 \\ 0 & 2 & 0 & -\sqrt{3} \\ 0 & 0 & \sqrt{3} & 0 \end{array} \right ),\nonumber\\
%&&J_z=\textrm{diag}\left(\frac{3}{2},\frac{1}{2},-\frac{1}{2},-\frac{3}{2} \right).
%\end{eqnarray*}

In units of $\mu_N=\frac{e}{2M}$ (M is proton mass), the magnetic moments of the proton and $\Delta^+$ as defined by eq.(\ref{magmom}) are:
\bea
&&\frac{\mu_p}{\mu_N}=1+\lambda_p=1+\frac{4M(m+M)}{3m^2}+\frac{2M^2}{3m^2}(F_1+F_2+F_3+F_5)~,\\
&&\frac{\mu_{\Delta^+}}{\mu_N}=\frac{M}{m}+(-\frac{1}{2}F_1+F_2)~.
\label{magmoment}
\eea
%\footnote{{when m is complex there seem to be no good solution to the definition of the magnetic moments.}}
%In the limit of m\to \infinty, proton anomalous magnetic moment \to 0(even with form factors?). Does it mean that the proton anomalous magnetic moment stems from the Delta baryon, and should I calculate loop correction to proton outline when calculating proton magnetic moment. The proton polarizability of the fundamental theory also comes from Delta baryon? Look at nucl-th/9610028 about polarizability and anomalous magnetic moment.
%Pion is the approximate goldstone for SU(2) (global) isospin sym breaking and W,Z are for SU(2) local weak sym breaking. Pion and W can both transit proton to neutron.
% The sign difference on the right hand side in eq.(\ref{magmom}) can be traced to the sign difference in the electric charge matrix for spin 1/2 and 3/2 components in eq.(\ref{elecharge}). 
When all the form factors are set to zero, $\mu_{\Delta^+}=\frac{e}{2m}$, so the g-factor of $\Delta^+$ is $\frac{2}{3}$, which agrees with expectations for that of an elementary spin 3/2 particle \cite{Belinfante:1953zz}. However, even in the minimally coupled theory,
the  spin 1/2 particle still has an anomalous magnetic moment due to the second term in the equation \eqref{magmoment}. Proton magnetic moment $\mu_p\simeq2.79$ is well measured and acts as a constraint on the form factors through eq.(\ref{magmoment}).  Intriguingly, the actual value is close to that of the minimally coupled theory, so that $F_1+F_2+F_3+F_5$ is approximately zero.

% the spin 1/2 particle in the fundament theory cannot be a point particle.

%use Compton scattering with photon lab energy around 350MeV, when the internal $\Delta^+$ approaches on-shell, to fit $\mu_{\Delta^+}$.
%fit pionphotonproduction processes: $p\gamma\to p\gamma\pi^0$
%For minimal electromagnetic interaction $\mu_{\Delta^+}=\frac{e}{2m}$ implies its g-factor to be $\frac{2}{3}$. 

$F_6$ does not contribute to the magnetic moments because it is higher order in the soft photon momentum k. $F_4$ does not enter $\mu_{\Delta^+}$ due to $\gamma_\rho \psi^\rho=0$ for spin 3/2 solution. Also, $F_4$ does not appear in the proton magnetic moment as can also be shown from the e.o.m. %, although not so manifest as the spin 3/2 case. 
%The e.o.m. $[p_\mu \Gamma^{\mu\lambda}{}_\rho-m\Theta^\lambda{}_\rho]\psi^\rho=0$ for spin 1/2 solution implies( is equivalent to) two simpler equations: $p^\nu \gamma_\rho \psi^\rho(p)=3M\psi^\nu(p)-M\gamma^\nu\gamma_\rho \psi^\rho(p)$, and $(\slash\!\!\!p-M)\psi^\rho=0$. Using these two simpler equations, then it is easy to prove that $F_4$ does not contribute to proton magnetic moment.

%to prove a complete list of vertexes using (simplified) e.o.m. still seems hard.

In the limit of degenerate mass for proton and $\Delta^+$ (where in reality the mass gap is indeed small: $\frac{|m-M|}{M}\sim 0.3$), we can calculate the transition amplitudes between slowly moving proton and $\Delta^+$. We take $\Delta^+$ at rest: $p_1=(m,0,0,0)$ and proton momentum $p_2=(\sqrt{M^2+k^2},0,0,k)$ and work in the degenerate $M \to m$ limit. We take the (virtual) photon to be left polarized $A_{\tiny{L}} = \frac{1}{\sqrt{2}}(0,1,i,0)$ or right polarized 
$A_{\tiny{R}}=\frac{1}{\sqrt{2}}(0,1,-i,0)$ and we calculate $e \, \bar{u}_4 \,(p_1,\sigma_1) \,\tilde{\Gamma}^\mu  \,u_2(p_2,\sigma_2)  \, A_{L/R\mu}$. For small $k$, the transition is $\mathcal O(k)$ and at first order in $k$ the result is: %\footnote{{G is replaced with $G/4/Sqrt[6]$, revise other places}}
\bea
&&e ~\bar{u}_4(p_1,\sigma_1)~\tilde{\Gamma}^\mu ~u_2(p_2,\sigma_2) ~A_{L\mu}=e \,\frac{2}{\sqrt{3}}\,G \, k ~ \left(\begin{array}{cc} \sqrt{3}/2& 0\\ 0 & 1/2 \\ 0& 0        \\0&0\end{array}\right)_{\sigma_1,\sigma_2} +\mathcal O(k^2),\nonumber\\
&&e ~\bar{u}_4(p_1,\sigma_1)~\tilde{\Gamma}^\mu ~u_2(p_2,\sigma_2) ~A_{R\mu}=e \,\frac{2}{\sqrt{3}}\,G \,k ~\left(\begin{array}{cc} 0&0    \\ 0&0 \\ -1/2 & 0\\ 0 & - \sqrt{3}/2\end{array}\right)_{\sigma_1,\sigma_2} +\mathcal O(k^2),
\label{Gdefinition}
\eea
where
%\footnote{for $e \bar{u}_2\tilde{\Gamma}^\mu u_4 e_{\mu}$, the combination appearing is $G^\dagger$.} 
$\tfrac{2}{\sqrt{3}}G=\frac{1}{6}\,(8 + 2 \, F_1+8 \, F_2+F_5 -i \,F_4)$ determines the magnetic transition amplitude between the proton and the $\Delta^+$. The entries of the matrix are the appropriate Clebsch-Gordan coefficients. $G$ is an important parameter and as we will see in the next Section makes the dominant contribution to the static magnetic polarizability in our model.
% G/ (sqrt(3)/2) is the QM normalized value

\section{Compton Scattering Cross Section and Polarizabilities}\label{sec:crosssec}

At tree level, the Feynman diagrams for proton Compton scattering are shown in Fig.\ref{fig2}. For s and u channels, the vertices were given in the previous section in eq.(\ref{genvertex}, \ref{nonminvertex}, \ref{antisymtensor}, \ref{sixthtensor}). 
%The proton wave function is:
%\beq
%(u_2){}^\rho(k,\sigma)=(LV^\rho{}_\lambda(k,M)\otimes LS(k,M))(u_2){}^\lambda(0,\sigma),
%\eeq
%where LV and LS are boost matrix for vector and dirac spinor fields, and $u_2(0,\sigma)$ are rest frame spin 1/2 solution:
%\bea
%&&u_2(0,+\frac{1}{2})=\frac{\sqrt{2M}}{2\sqrt{3}(3z-1)}(0, 0, 0, 0, 0,1, 0, -1, 0, i, 0, -i, 1, 0, -1, 0)^T,\nonumber \\
%&&u_2(0, -\frac{1}{2})=\frac{\sqrt{2M}}{2\sqrt{3}(3z-1)}(0,0, 0, 0, 1, 0, -1, 0, -i, 0, i, 0, 0, -1, 0, 1)^T.
%\eea
%The wave functions are written in 16 component form, where the first four components correspond to $\rho=0$ in $u_2{}^\rho$, the second four components correspond to $\rho=1$ in $u_2{}^\rho$, etc.. The vertex and propagator are dealt with as $16\times16$ matrixes similarly.\par

For pion exchange t-channel diagram,
%\footnote{the $\sigma$-exchange we do not consider here since it does not help the fitting very much.}
 there is no contribution from $\Delta^+$, and we use the familiar Dirac spinor for proton wave function. 
%To be strict, we should have proposed the coupling of pion to proton and $\Delta^+$ in this new theory for proton-$\Delta^+$ system, but that would introduce more model dependence. Also we believe the new interaction for pion and proton will not make much difference to this work since in the low energy limit the new and old interactions should give the same prediction, not to mention that pion channel contribution to the amplitude is small itself. 
The relevant interaction Lagrangian is:
\beq
{\cal L}_{\mbox{\tiny int}}=i \, g_\pi \, \bar u\,  \gamma^5 \, u \, \pi^0 + \frac{1}{8}\, F_{\pi\gamma\gamma} \, \epsilon_{\mu\nu\rho\lambda}\, F^{\mu\nu}\, F^{\rho\lambda} \, \pi^0.
\eeq

%about the sign in this vertex and that in Pascalutsa's paper: if Pascalutsa used the same convention as in Peskin's book: \epsilon^{0123}=+1 and \gamma_5=diag{-1,-1,+1,+1}, it seems that my definition of the sign of the vertex is minus to Pascalutsa's definition, this I should pay attention to. by the way my \gamma_5 is actually -i times the above one.

The incoming proton and photon have 4-momentum $k_1$ and $k_2$, and the outgoing proton and photon $k_4$ and $k_3$ respectively. With the above Feynman rules, the tree level amplitude is the sum of the three diagrams:
\bea
&&{\mathcal A}_{\sigma_1,\sigma_4,\lambda_2,\lambda_3}\nonumber\\
 =&&A^{\mu\nu}\epsilon_\mu(k_2, \lambda_2)\ \epsilon_\nu^*(k_3,\lambda_3)\nonumber\\
=&& \Big[(ie)^2 \ \ {\bar u_2}(k_4,\sigma_4) \ {\tilde \Gamma}^\mu \ (-i)S(k_1-k_3)\ {\tilde \Gamma}^\nu \ u_2(k_1,\sigma_1)\nonumber\\
&&+ (ie)^2\ \ {\bar u_2}(k_4,\sigma_4) \ {\tilde \Gamma}^\nu \ (-i)S(k_1+k_2)\ {\tilde \Gamma}^\mu \ u_2(k_1,\sigma_1)\nonumber\\
&&+ {\scriptstyle \frac{i g_\pi \, F_{\pi\gamma\gamma}}{(k_1-k_4)^2 - m_\pi^2}}\,{\bar u}(k_4,\sigma_4) \,  \gamma^5 \, u(k_1,\sigma_1) \, \epsilon^{\mu\nu\sigma\rho} \, k_{2\sigma} \, k_{3\rho}\Big]\nonumber\\
&&\epsilon_\mu(k_2, \lambda_2)\ \epsilon_\nu^*(k_3,\lambda_3).
\label{matrixele}
\eea

\begin{figure}[htdp]
\setlength{\unitlength}{1cm}
\center{\includegraphics[height=7cm,width=11cm]{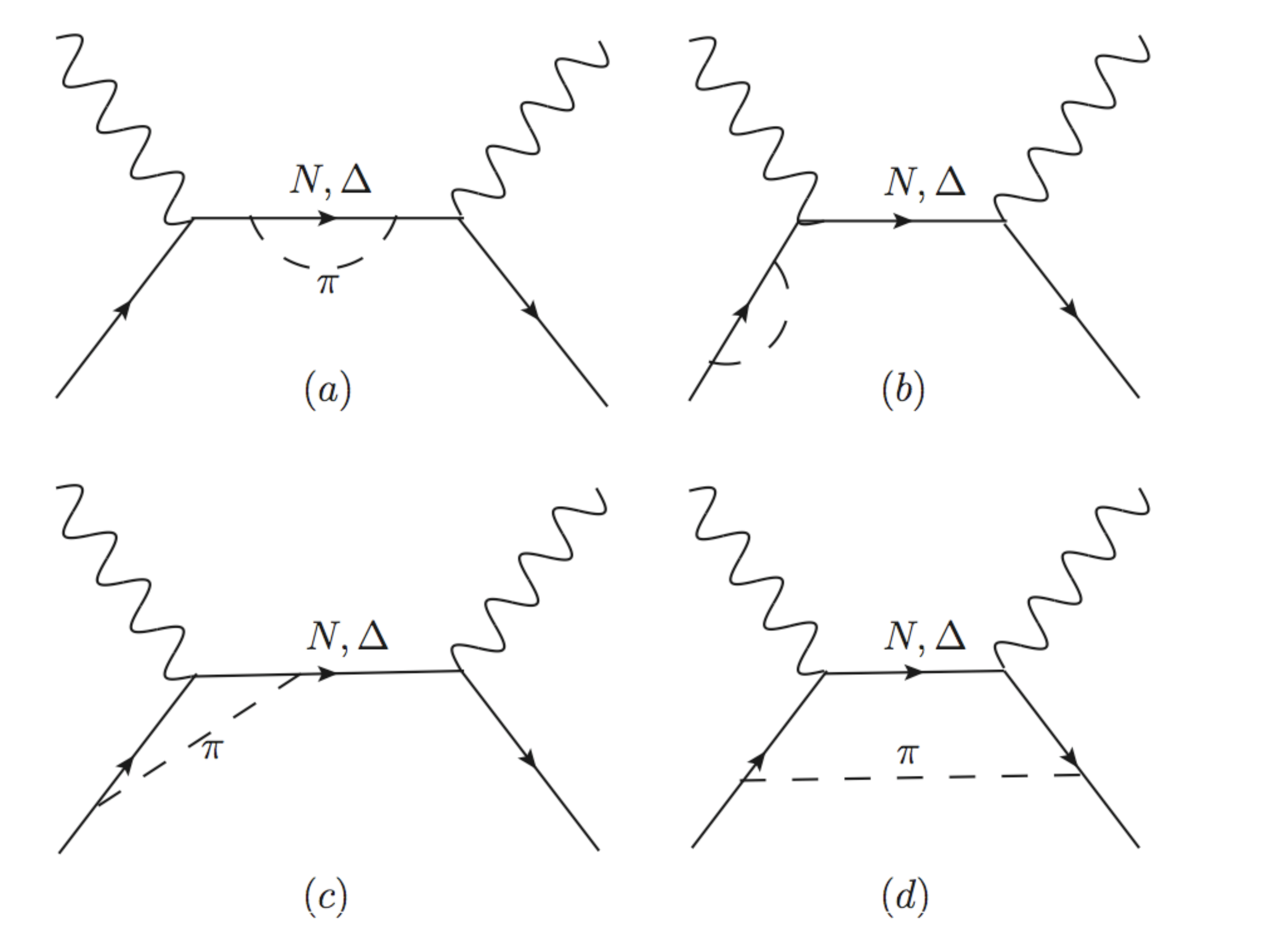}}
\caption{The one-loop level  pion corrections: (a) and (b) are the self-energy diagrams; (c) and (d) are pion vertex corrections. More diagrams emerge at higher loop levels.}
\label{picture2}
\end{figure}

In addition to diagrams in Fig.\ref{fig2}, strong interactions contribute through the pion one-loop diagrams as in Fig. \ref{picture2}. Above the pion-production threshold, the diagram a) contributes to the  imaginary part of the self-energy of $\Delta^+$ and determines the line-shape of the resonance.  In principle, the imaginary part depends on c.m. momentum squared s, and all these diagrams should be taken into account at one-loop order \cite{McGovern:2012ew}. 
For our purposes, we make an estimate of this effect by setting the imaginary part of $\Delta^+$ mass m to the observed width at the resonance, i.e. we analytically extend the above matrix element by substituting m with $(m_0-i\frac{\Gamma}{2})\sim (1210-50i)$MeV everywhere it appears in the matrix element. % \footnote{Notice that the proton wave function is also affected through the dependence of z on m: $z=\frac{m+2M}{6M}$, although proton is a stable particle, while $\Delta^+$ wave function not altered.}. 
Despite modification of both the vertex and the propagator, this procedure preserves the Ward identity due to the analyticity of  eq. (\ref{wardid}) in $m$. %, so that $k_{2\mu} \, A^{\mu\nu}=0$, $k_{3\nu} \, A^{\mu\nu}=0$.% ????? Also, the electric charge in eq.(\ref{elecharge}) does not change under this substitution. 

It is verified that our result satisfies Low's theorem \cite{Low1954}, namely that for the low energy Compton scattering on spin 1/2 particles, the amplitudes expanded to first order of photon energy are completely determined by the mass, electric charge and magnetic moment of the spin 1/2 particle. 
%Although the theorem was proved in case of Dirac spinor, it nevertheless applies to the current case. 
According to the theorem, in lab frame with photon incident energy $\omega$, the amplitudes are:
\bea
&&{\mathcal A}_{\sigma_1,\sigma_4,\lambda_2,\lambda_3}\nonumber\\
=&&\frac{e^2}{M} \vec{\epsilon}_{\lambda_2}\cdot \vec{\epsilon}^*_{\lambda_3}\delta_{\sigma_1 \sigma_2}-\frac{i e \omega}{M}(2 \mu_p-\frac{e}{2M})( \vec{\epsilon}^*_{\lambda_3}\times \vec{\epsilon}_{\lambda_2})\cdot \vec{\sigma}_{\sigma_1\sigma_2}\nonumber\\
&&+\frac{ie\mu_p}{\omega M}(\vec{\epsilon}_{\lambda_2}\cdot \vec{k}_3(\vec{\epsilon}^*_{\lambda_3}\times \vec{k}_3)-\vec{\epsilon}^*_{\lambda_3}\cdot \vec{k}_2(\vec{\epsilon}_{\lambda_2}\times \vec{k}_2))\cdot \vec{\sigma}_{\sigma_1\sigma_2}\nonumber\\
&&+\frac{2i\mu_p^2}{\omega}((\vec{\epsilon}^*_{\lambda_3}\times \vec{k}_3)\times(\vec{\epsilon}_{\lambda_2}\times \vec{k}_2))\cdot \vec{\sigma}_{\sigma_1\sigma_2}+O(\omega^2).
\label{lowtheorem}
\eea

Expansion of the cross section to second order in photon energy $\omega$ is exactly in the form of eq.(\ref{polarizability}) with $\bar\alpha \pm \bar\beta$: 
\bea
\bar\alpha+\bar\beta\sim && -0.860-0.556 F_1-1.789 F_2-0.240 F_5\nonumber\\
&&-0.069 F_1^2-0.961 F_2^2-0.015 F_4^2-0.020 F_5^2\nonumber\\
&&-0.536 F_1 F_2-0.023 F_1 F_3-0.085 F_1 F_5+0.009 F_2 F_3\nonumber\\
&&-0.234 F_2 F_5-0.007 F_3 F_5+\alpha_B+\beta_B\ (10^{-4} fm^3),\nonumber\\
&&\nonumber\\
\bar\alpha-\bar\beta\sim &&1.894+1.284 F_1+2.602 F_2+0.202 F_3+0.650 F_5\nonumber\\
&&+0.146 F_1^2+1.339 F_2^2+0.028 F_3^2+0.023 F_4^2+0.064 F_5^2\nonumber\\
&&+0.728 F_1 F_2+0.069 F_1 F_3+0.177 F_1 F_5+0.101 F_2 F_3\nonumber\\
&&+0.447 F_2 F_5+0.067 F_3 F_5+\alpha_B-\beta_B\ (10^{-4} fm^3).
\label{numpolarizability}
\eea
where $\alpha_B$ and $\beta_B$ are bare polarizabilities defined in eq.(\ref{barepol}). 
%Substitute $m=1232MeV$ and $M=938MeV$ in eq.(\ref{polarizabilityexpressnew}) we have approximately:

%Substituting the complex value of m in the amplitudes we get:
%\bea
%\bar\alpha+\bar\beta\sim &&-0.0734 F_1^2-0.570 F_2 F_1-0.024 F_3 F_1-0.090 F_5 F_1\nonumber\\
%&&-0.590 F_1-1.024 F_2^2-0.016 F_4^2-0.022 F_5^2\nonumber\\
%&&-1.909 F_2+0.009 F_2 F_3-0.249 F_2 F_5-0.007 F_3 F_5\nonumber\\
%&&-0.256 F_5-0.917+\alpha_B+\beta_B\ (10^{-4} fm^3),\nonumber\\
%&&\nonumber\\
%\bar\alpha-\bar\beta\sim && 0.155 F_1^2+0.772 F_2 F_1+0.073 F_3 F_1+0.187 F_5 F_1\nonumber\\
%&&+1.350 F_1+1.418 F_2^2+0.030 F_3^2+0.024F_4^2\nonumber\\
%&&+0.067 F_5^2+2.764F_2+0.107 F_2 F_3+0.213 F_3\nonumber\\
%&&+0.470 F_2 F_5+0.071 F_3 F_5+0.686 F_5+1.998+\alpha_B-\beta_B\ (10^{-4} fm^3).
%\eea
% \footnote{The minimal interaction, that is when all the form factors $F_1,\cdots,F_6$ and $\alpha_B,\beta_B$ are 0, already induces non-zero polarizabilities. }
%think why \bar\alpha+\bar\beta is negative for bare theory.

%F3 contribution seems very small in \bar\alpha+\bar\beta

\subsection{Amplitudes at the $\Delta^+$ Pole}

In reality, $\Delta^+$ and proton have small mass gap:
\beq
m=M (1+x- i \, y),
\eeq
where $x\sim 0.3$ and $y\sim 0.05\sim \frac{1}{2} \, x^2$. And around the $\Delta^+$ resonance, the photon momentum q is of the same order as x, we can approximate the amplitudes to lowest non-trivial order of q, x and y.

First, at the $\Delta^+$ pole position, the contribution to the pole mainly comes from the first term in the propagator in s-channel where the momentum propagated is $k_1+k_2$. In this case, the denominator contributing to the pole is $(k_1+k_2)^2-m^2=(E_{CM}-m)(E_{CM}+m)$. We multiply the amplitudes by $(E_{CM}-m)$ and then expand it with respect to q, x, and y. We define $q= r_1 x$ and $y=r_2 x^2$. Around the peak, $r_1\sim 1$ and $r_2 \sim 0.5$. Then we can  expand the amplitudes multiplied with $(E_{CM}-m)$ with respect to x, and finally set $r_1=1$(at the peak) and $r_2=\frac{y}{x^2}$. In center of mass frame, we rotate the proton wave functions to make it polarized along its direction of moving. That is, for proton moving in direction $\theta$ with respect to z-axis:
\bea
&&\tilde{u}_2(p,\frac{1}{2}) =\cos\Big(\frac{\theta}{2}\Big) ~u_2(p,\frac{1}{2})+\sin\Big(\frac{\theta}{2}\Big) ~u_2(p,-\frac{1}{2}),\nonumber\\
&&\tilde{u}_2(p,-\frac{1}{2}) =-\sin\Big(\frac{\theta}{2}\Big) ~u_2(p,\frac{1}{2})+\cos\Big(\frac{\theta}{2}\Big) ~u_2(p,-\frac{1}{2}).
\label{polarizedproton}
\eea
Then the approximate amplitudes are:
\bea
&&{\cal A}_{ppRR}=\frac{(2 i r_2+1 -2|G|^2) x^2\cos^3\frac{\theta}{2}}{2(E_{CM}-m)},\nonumber\\
&&{\cal A}_{ppRL}=\frac{(-2 i r_2-1 -2|G|^2) x^2\cos\frac{\theta}{2}\sin^2\frac{\theta}{2}}{2(E_{CM}-m)},\nonumber\\
&&{\cal A}_{ppLR}=\frac{(-2 i r_2-1 -2|G|^2) x^2\cos\frac{\theta}{2}\sin^2\frac{\theta}{2}}{2(E_{CM}-m)},\nonumber\\
&&{\cal A}_{ppLL}=\frac{(6 i r_2+3 +2|G|^2+3(2 i r_2+1-2|G|^2)\cos\theta) x^2\cos\frac{\theta}{2}}{12(E_{CM}-m)},\nonumber\\
&&{\cal A}_{pmRR}=\frac{(2 i r_2+1 -2|G|^2) x^2\cos^2\frac{\theta}{2}\sin\frac{\theta}{2}}{2(E_{CM}-m)},\nonumber\\
&&{\cal A}_{pmRL}=\frac{(-2 i r_2-1 -2 |G|^2) x^2\sin^3\frac{\theta}{2}}{2(E_{CM}-m)},\nonumber\\
&&{\cal A}_{pmLR}=\frac{(-6 i r_2-3 +2|G|^2+3(2 i r_2+1+2|G|^2)\cos\theta) x^2\sin\frac{\theta}{2}}{12(E_{CM}-m)},\nonumber\\
&&{\cal A}_{pmLL}=\frac{(2 i r_2+1 -2|G|^2) x^2\cos^2\frac{\theta}{2}\sin\frac{\theta}{2}}{2(E_{CM}-m)}.
\label{peakbehavior}
\eea
In the above expressions, $x=\frac{\mbox{\tiny Re}(m)-M}{M}$, $r_2=\frac{y}{x^2}=-\frac{\mbox{\tiny Im}(m)}{M x^2}$. $G$ is the combination of form factors which describes the $\gamma N \Delta^+$ transition amplitude as described in the previous section in eq.(\ref{Gdefinition}). For the subscript of the amplitudes, the first/second p/m stands for the final/initial proton polarization and the first/second L/R for final/initial photon polarization. Here only 8 of the 16 amplitudes are given, since the other 8 are related by parity:
\bea
&&{\cal A}_{mmRR}={\cal A}_{ppLL},\ \ \ \ \ \ \ {\cal A}_{mpRR}=-{\cal A}_{pmLL},\nonumber\\
&&{\cal A}_{mmRL}={\cal A}_{ppLR},\ \ \ \ \ \ \ {\cal A}_{mpRL}=-{\cal A}_{pmLR},\nonumber\\
&&{\cal A}_{mmLR}={\cal A}_{ppRL},\ \ \ \ \ \ \ {\cal A}_{mpLR}=-{\cal A}_{pmRL},\nonumber\\
&&{\cal A}_{mmLL}={\cal A}_{ppRR},\ \ \ \ \ \ \ {\cal A}_{mpLL}=-{\cal A}_{pmRR}.
\eea

In the same limit, the magnetic polarizability has an approximation:
\beq
\bar\beta=-\frac{4\alpha |G|^2}{3 x M^3}+\beta_B.
\eeq
This may imply that the magnetic polarizability has something to do with proton-$\Delta$ (magnetic) transition.

\section{Fitting Data}
\label{sec:fit}
% \footnote{cannot download "low energy $\gamma p$ scattering and determination of proton polarizabilities" by P.S.Baranov 2001 and "Experimental status of the electric and magnetic polarizabilities of a proton" by P.S.Baranov 2001.}

We fit the model to the 714 proton Compton scattering datapoints from 8 experiments\cite{Hallin1993,Baranov1974,Zieger92,Hunger97,Blanpied01,MAMI2001,Olmos2001,MacGibbon1995}. Only data points with photon incident energy smaller 455MeV are used, in the  so-called first resonance region. In principle, one can also compare the model predictions with polarized measurements, where some data is available \cite{al:1993aa, Wada:1981ab}.

For several reasons, we set $F_4=0$ in our fitting. First, $F_4$ does not enter in the expressions of proton and $\Delta^+$ magnetic moments eq.(\ref{magmoment}). Second, in all fits we attempted, the best fit value of  $F_4$ was nearly exactly zero, and in any case statistically consistent with zero.

The parameters we use to fit are chosen to be $F_1$, $\mu\equiv\frac{\mu_\Delta^+}{\mu_N}=\frac{M}{m}+F_2-\frac{1}{2}F_1$, $G=\frac{1}{4\sqrt{6}}(2F_1+8F_2+F_5+8)$, $F_6$ and the bare polarizabilities $\alpha_B$ and $\beta_B$ in eq.(\ref{poleffLag}). $F_3$ is constrained using proton magnetic moment, see eq.(\ref{magmoment}).

We minimize
%\footnote{Is this $\chi^2$ suitable. In Baranov 2000 paper, they use statistical error for each data point and systematic error to rescale each experiment.} 
$\chi^2=\sum_{i=1}^{714}\frac{((\frac{d\sigma}{d\Omega})_i^{\mbox{\tiny\it calc}}-(\frac{d\sigma}{d\Omega})_i^{\mbox{\tiny\it data}})^2}{\sigma_{\mbox{\tiny (stat)}i}^2+\sigma_{\mbox{\tiny (syst)}i}^2}$.
We did not attempt to rescale the data of each experiment within its own systematical uncertainty to see if it would lead to better consistency between datasets as it was done in  \cite{Baranov:2000na}. The optimal set of parameters was found to be: $F_1=-27.5$, $\mu=14.2$, $G =3.13$, $F_6 = 12.9$, $\alpha_B = 7.5$, $\beta_B = -8.2$ with $\chi^2\sim 6.3\times 10^3$.
%\footnote{the coupling constant of pion channel may need some adjusting and the form factors of the pion couplings are not considered. It seems that when considering pion form factors and no sigma exchange, the fit becomes much worse.}

\begin{figure}[htdp]
\setlength{\unitlength}{1cm}
\includegraphics[height=4cm,width=6cm]{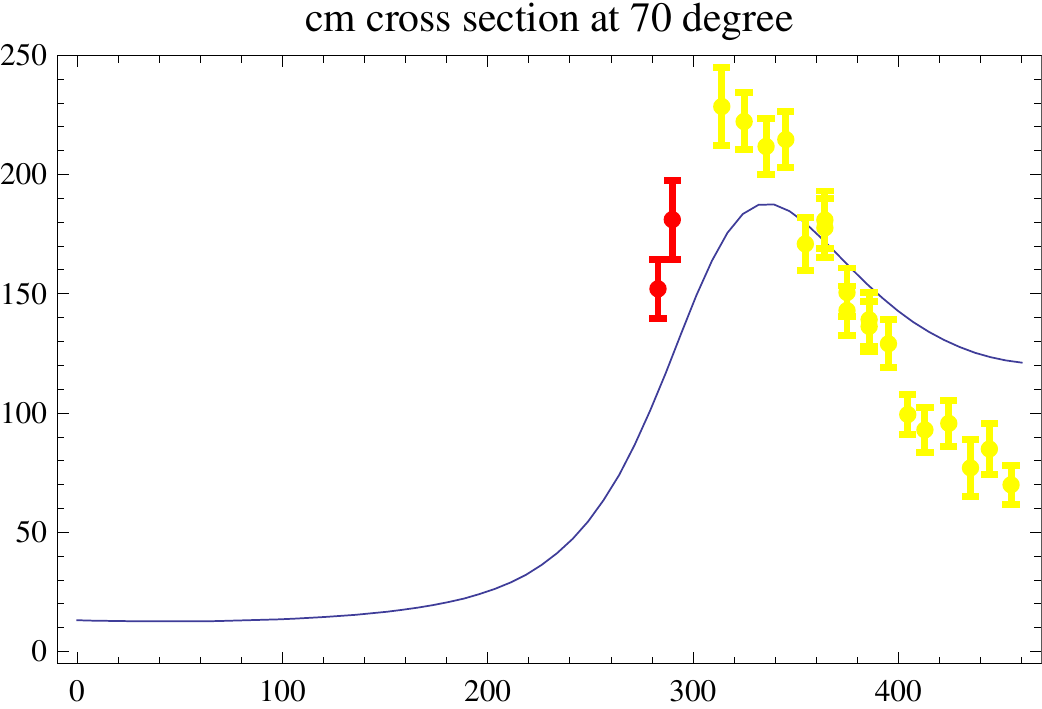}
\includegraphics[height=4cm,width=6cm]{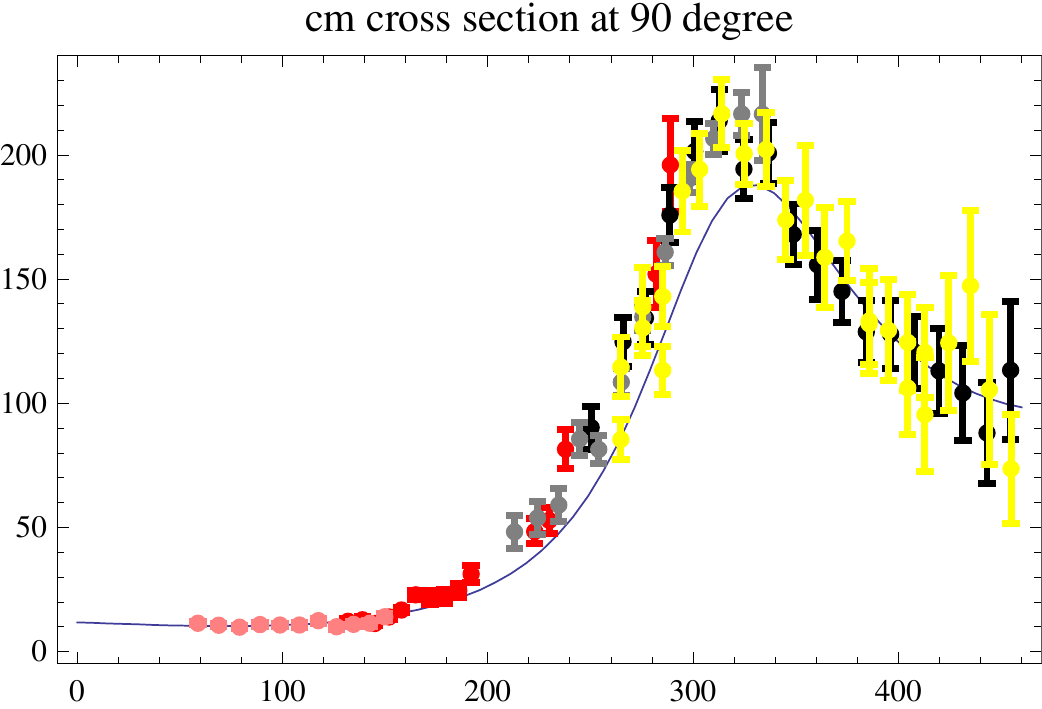}\\
\includegraphics[height=4cm,width=6cm]{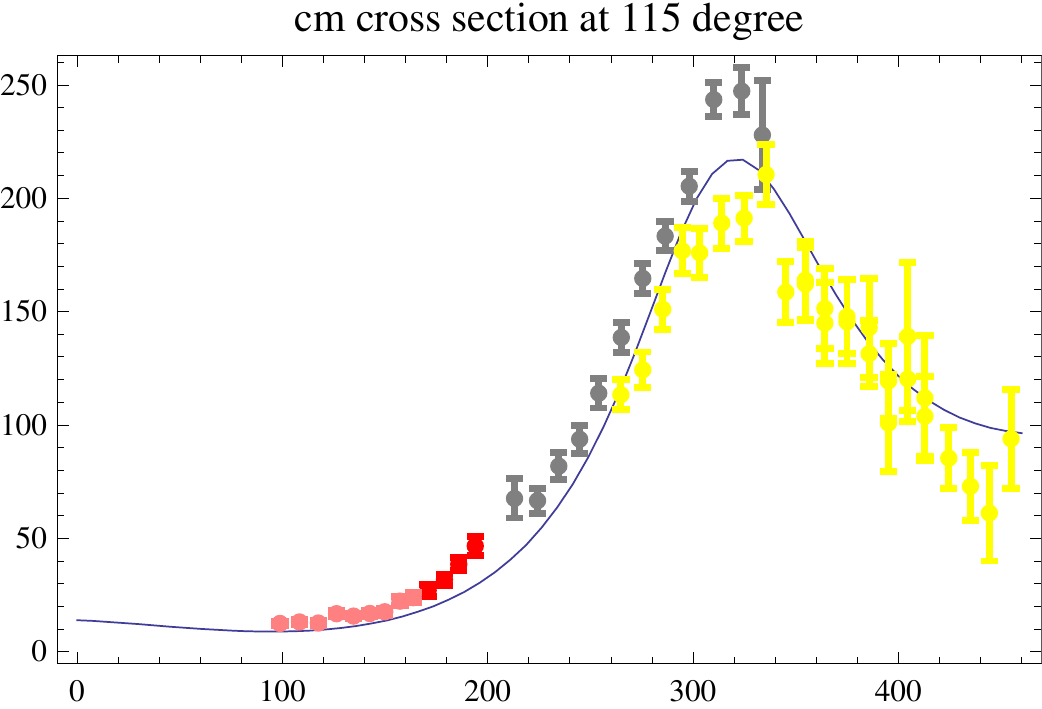}
\includegraphics[height=4cm,width=6cm]{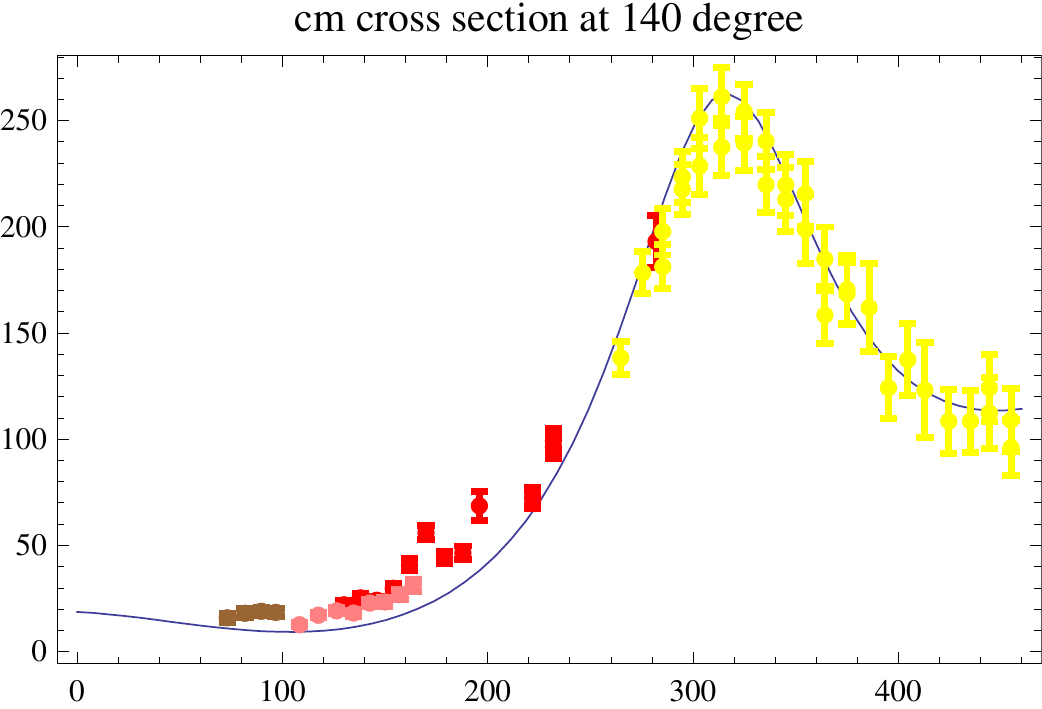}
\caption{Fixed c.m. angle cross section and the data points, where the parameters from the fitting to all the 714 data points are used for the theoretical cross section curve. The x-axis is lab frame photon energy and y-axis is c.m. frame differential cross section in unit of nb. %Those data measured or recorded in lab frame have been converted to c.m. frame using the transition in eq.(\ref{labcmrela}). 
For \cite{Hallin1993,Baranov1974,Zieger92,Hunger97,Blanpied01,MAMI2001,Olmos2001,MacGibbon1995}, we use colors: Green, Blue, Black, Brown, Red, Gray, Pink and Yellow respectively. The angles of the data points included may differ from the values claimed by at most 3 degree.}
\label{angleplot}
\end{figure}

\begin{figure}[htdp]
\setlength{\unitlength}{1cm}
\includegraphics[height=4cm,width=6cm]{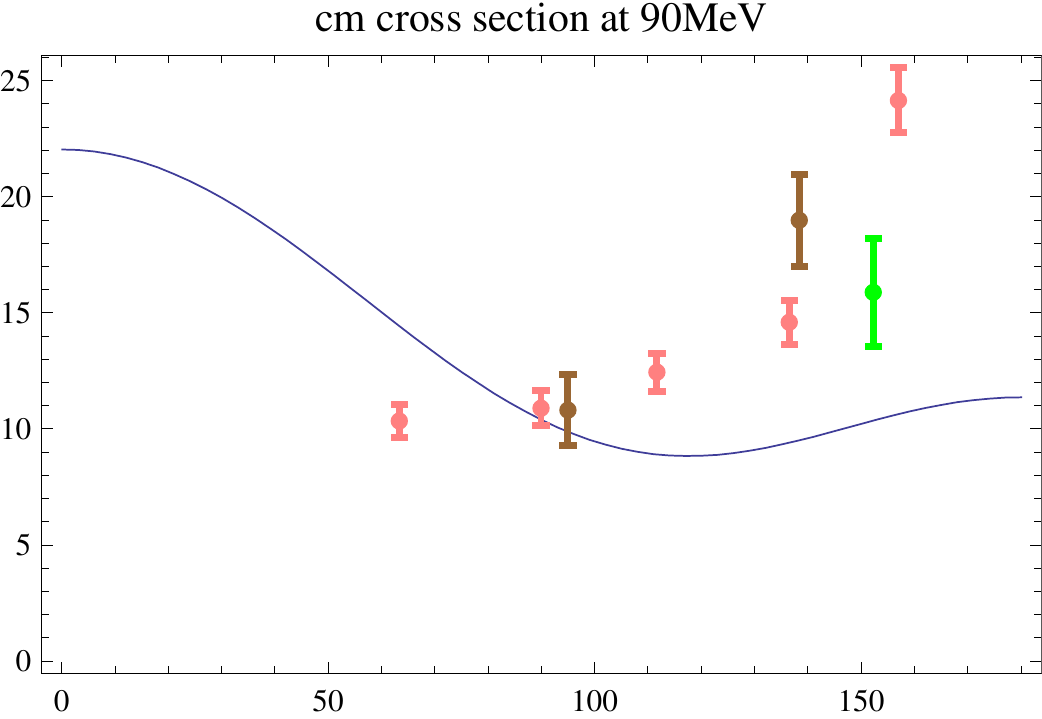}
\includegraphics[height=4cm,width=6cm]{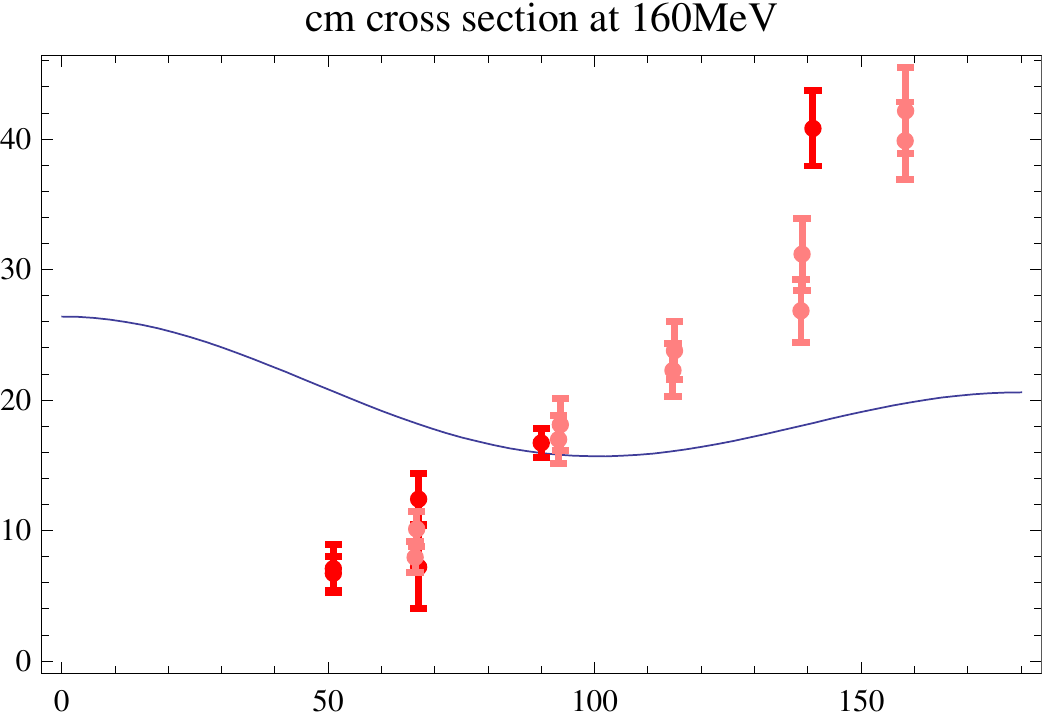}\\
\includegraphics[height=4cm,width=6cm]{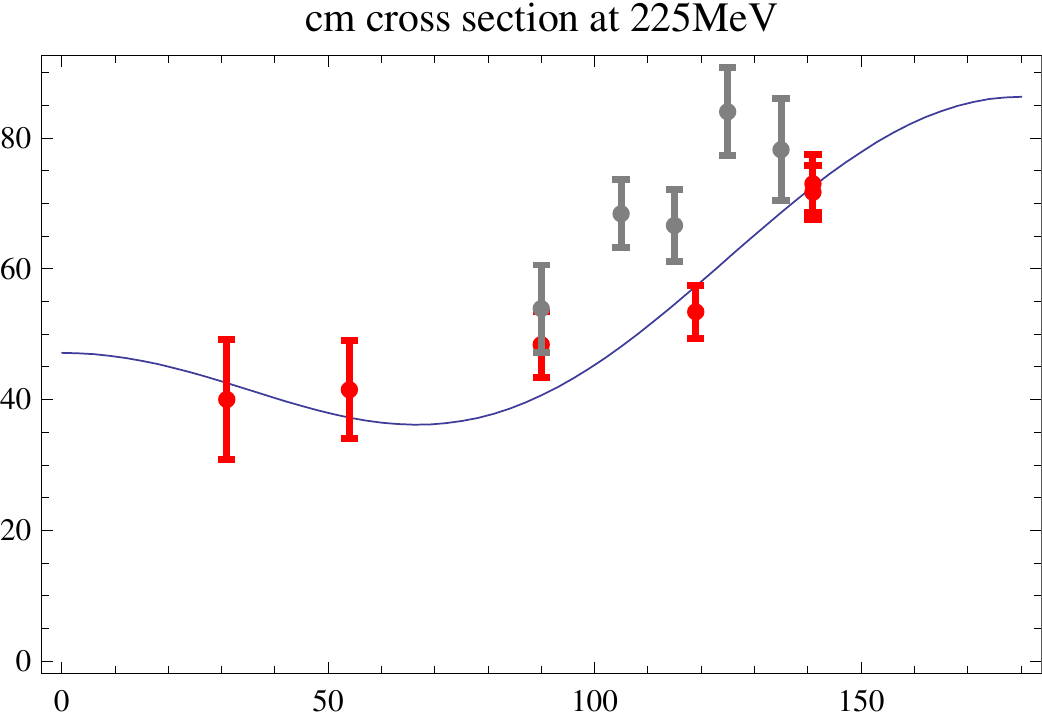}
\includegraphics[height=4cm,width=6cm]{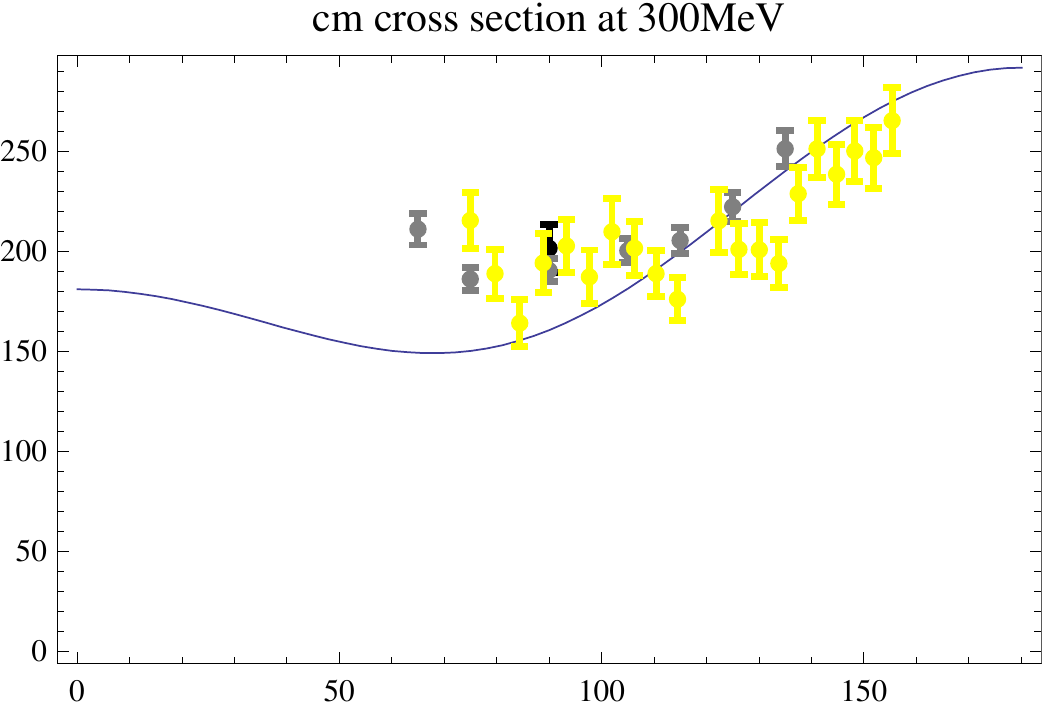}\\
\includegraphics[height=4cm,width=6cm]{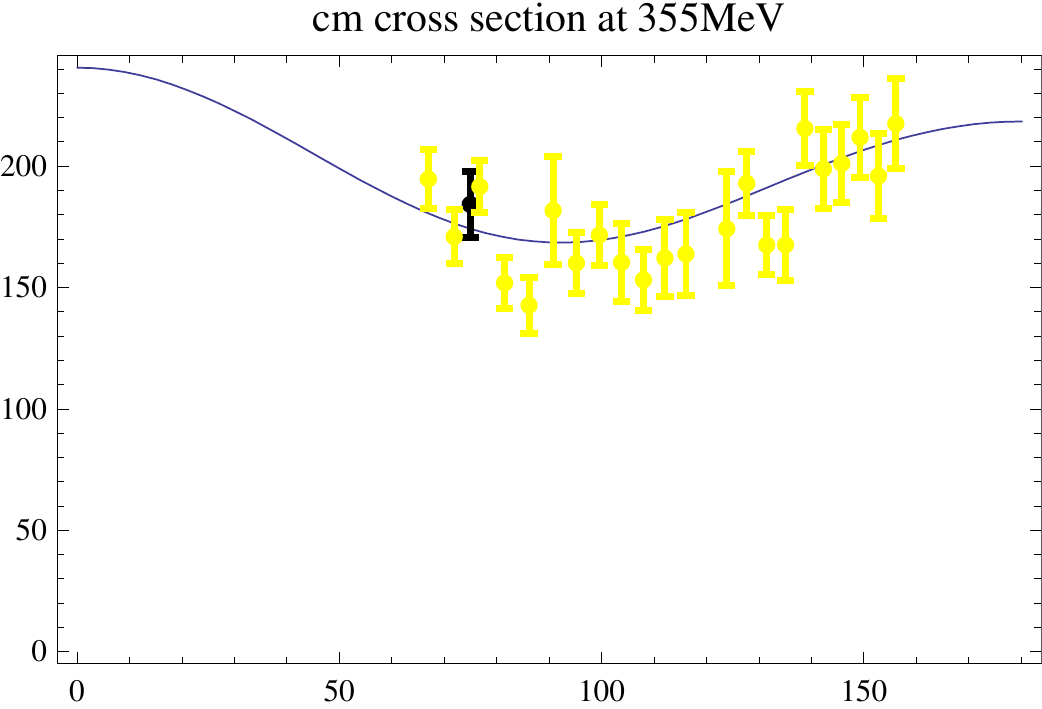}
\includegraphics[height=4cm,width=6cm]{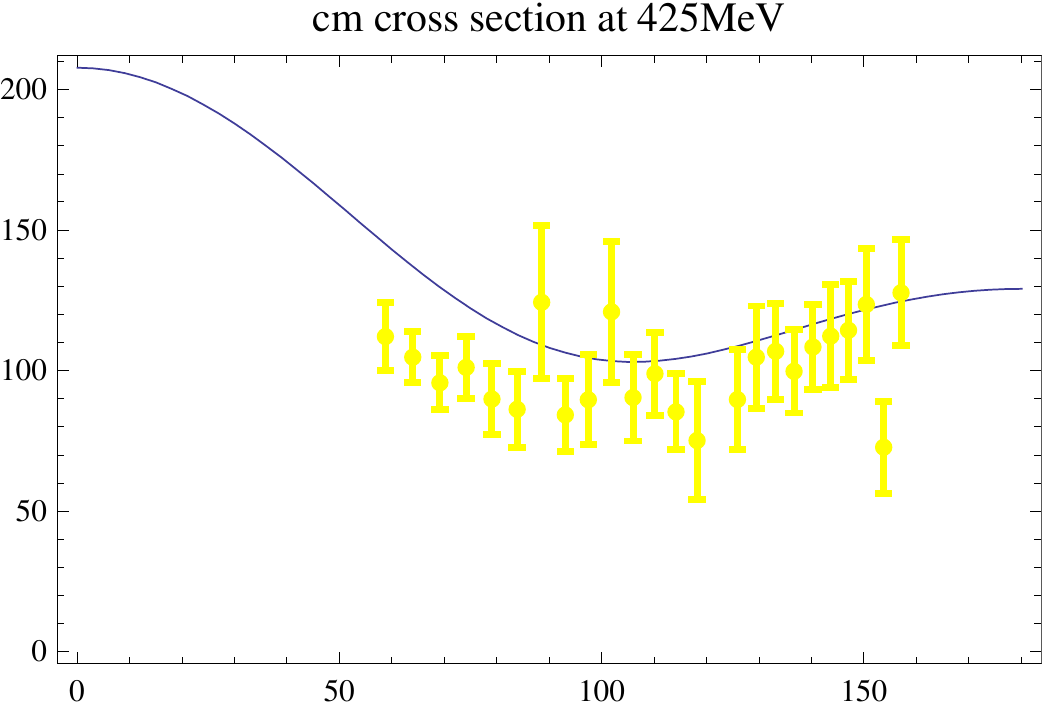}
\caption{Fixed c.m. angle cross section and the data points, where the parameters from the fitting to all the 714 data points are used for the theoretical cross section curve. The x-axis is c.m. frame scattering angle and y-axis is c.m. frame differential cross section in unit of nb. For \cite{Hallin1993,Baranov1974,Zieger92,Hunger97,Blanpied01,MAMI2001,Olmos2001,MacGibbon1995}, we use colors: Green, Blue, Black, Brown, Red, Gray, Pink and Yellow respectively. The incident photon energy of the data points included may differ from the values claimed by at most 4MeV.}
\label{energyplot}
\end{figure}

We plot the c.m. frame cross section using the above fit parameters, together with data points in Fig.\ref{angleplot} and \ref{energyplot}. Those data measured or recorded in lab frame have been converted to c.m. frame. From Fig.\ref{energyplot} it is seen that the low energy cross section fit badly.

On average for each data point the fit is of $3\sigma$ deviation from the experiment value. The $\Delta^+$ resonance region is fitted well, while the low energy cross section deviates greatly from data points. In fact, the 68 data points with incident photon energy smaller than 140MeV out of the total 714 data points contributes nearly a third of the total $\chi^2$. Our fit cannot take care of the low energy ($\lae$140MeV) data points and "high" energy ($\gae$200MeV) data points simultaneously. When giving a good fit in the resonance region, where most data points used in this paper lie in, the predicted cross section at low energy cannot account of the large asymmetry of the cross section data at forward and backward angles. Our fit cross sections at low energies are much higher than the data at forward angles and lower at backward angles. Since the polarizabilities are extracted according to low energy expansion of the cross section in eq.(\ref{polarizability}), it is expected that the predicted $\bar\alpha+\bar\beta$, calculated using eq.(\ref{numpolarizability}) where $F_2$ and $F_5$ are solved from the definitions of $\mu$ and $G$, is smaller than the experimental value, and $\bar\alpha-\bar\beta$ larger than experimental value. For the above fit values, $\bar\alpha+\bar\beta=-2.6 (10^{-4}fm^3)$, $\bar\alpha-\bar\beta=45.1(10^{-4}fm^3)$. %(The form factors contribute -1.2 and 31.5 to them respectively). 
By contrast, the original experiments have quoted values extracted from the same data of $\bar\alpha+\bar\beta$ at about $14.0 (10^{-4}fm^3)$, and $\bar\alpha-\bar\beta$ at about $10.0 (10^{-4}fm^3)$.

%rescaling of the experiments are not done here
We plot the c.m. frame cross section using the above fit parameters, together with data points in Fig.\ref{angleplot} and \ref{energyplot}. %Those data measured or recorded in lab frame have been converted to c.m. frame using the transition in eq.(\ref{labcmrela}).

%The difference between the predicted decay width and experimental value may have several explanations. First, as mentioned almost all of the experiments used to extract the decay amplitudes are $p+\gamma \to \pi+\gamma$ scattering yet we are using Compton scattering here. Then, in our fitting we did not rescale the experiments and if we had done a rescaling such that the data points around the peak were enhanced, the decay width extracted would increase too, as the same coupling $G$ is governing the peak in Compton scattering and decay width shown in eq.(\ref{peakbehavior}) and eq.(\ref{widthapprox}). Also, this fitting shows obvious deviation from data points, and especially in the pole range, the predicted cross sections are mostly below the data points, causing a smaller prediction for decay width, with the same reason just above.

\begin{figure}[htdp]
\setlength{\unitlength}{1cm}
$\begin{array}{ccccc}
&F_1&G&\mu&F_6\\
\raisebox{1.5cm}{$F_1$}&\raisebox{1.5cm}{$\not$}&\mbox{}\includegraphics[height=3cm,width=3cm]{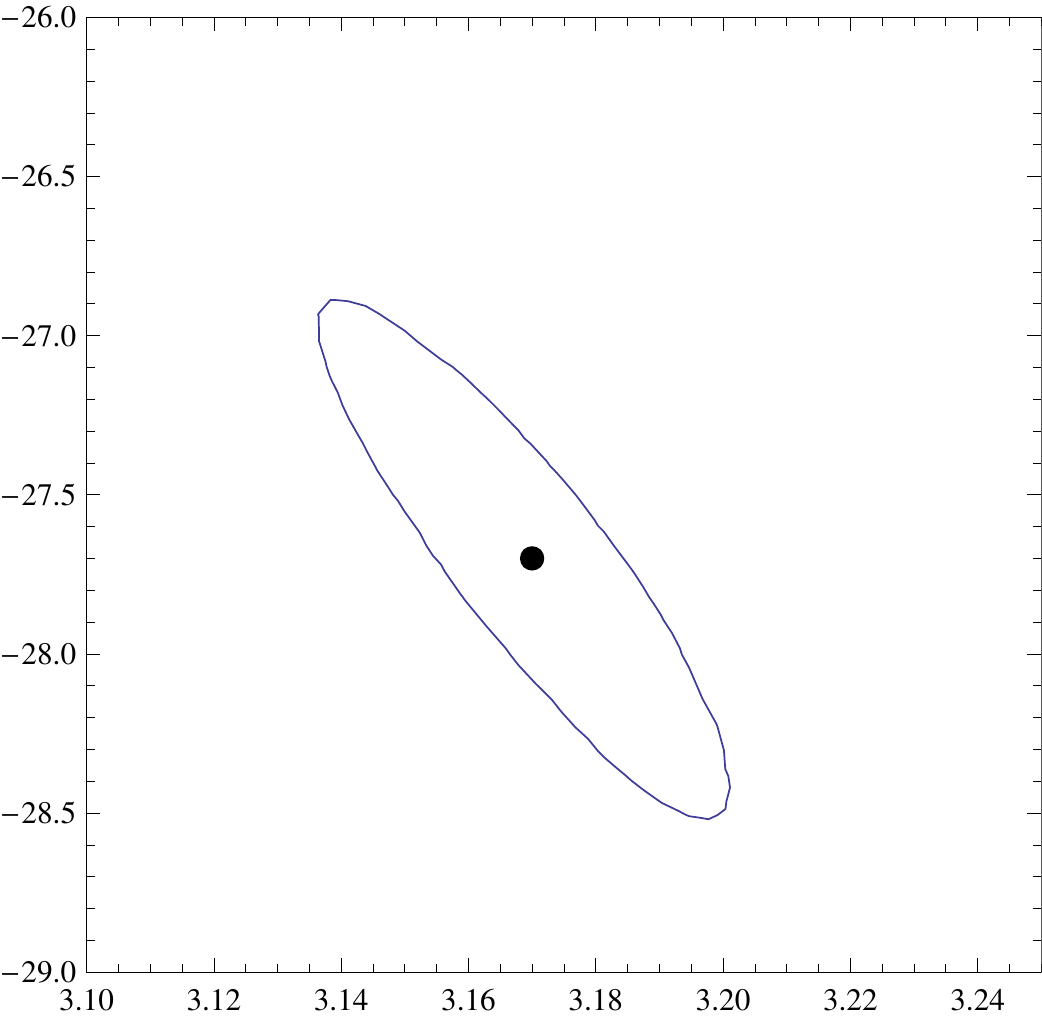}&\includegraphics[height=3cm,width=3cm]{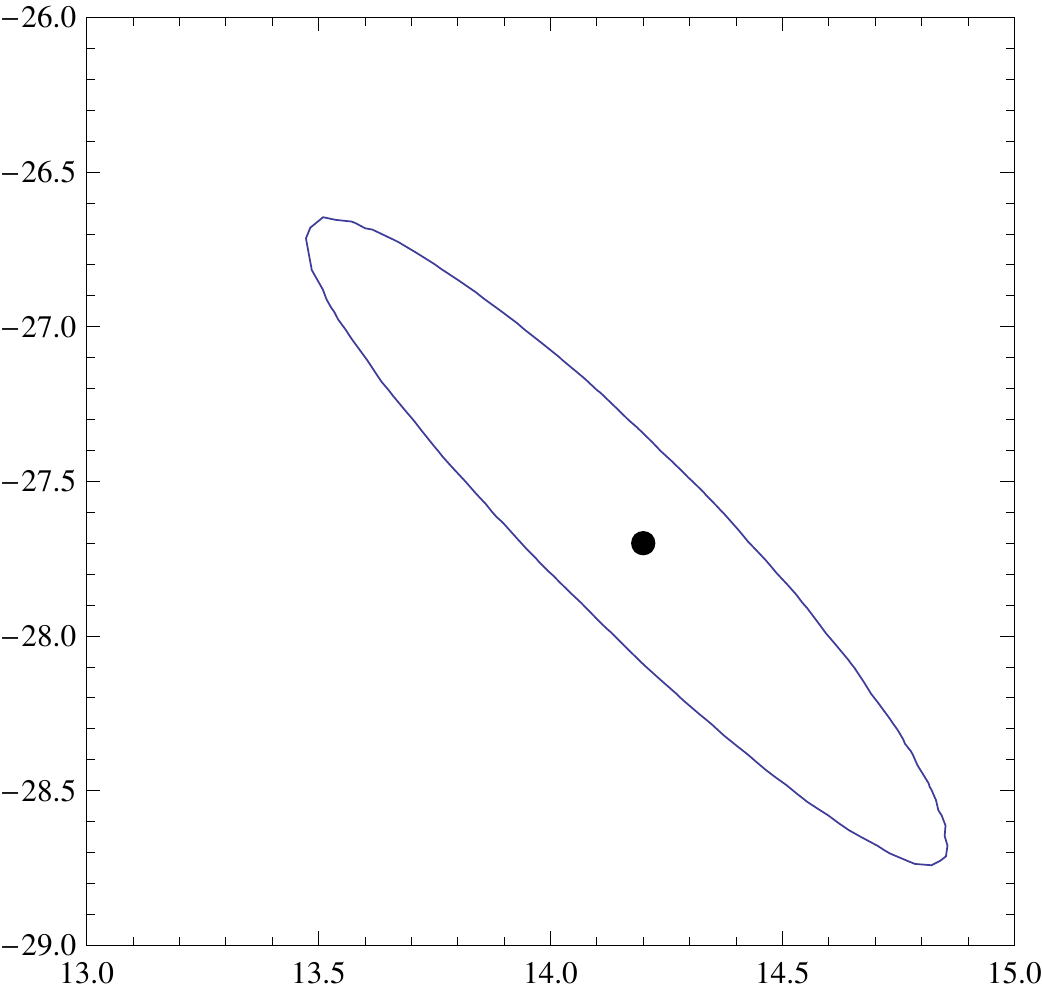}&\includegraphics[height=3cm,width=3cm]{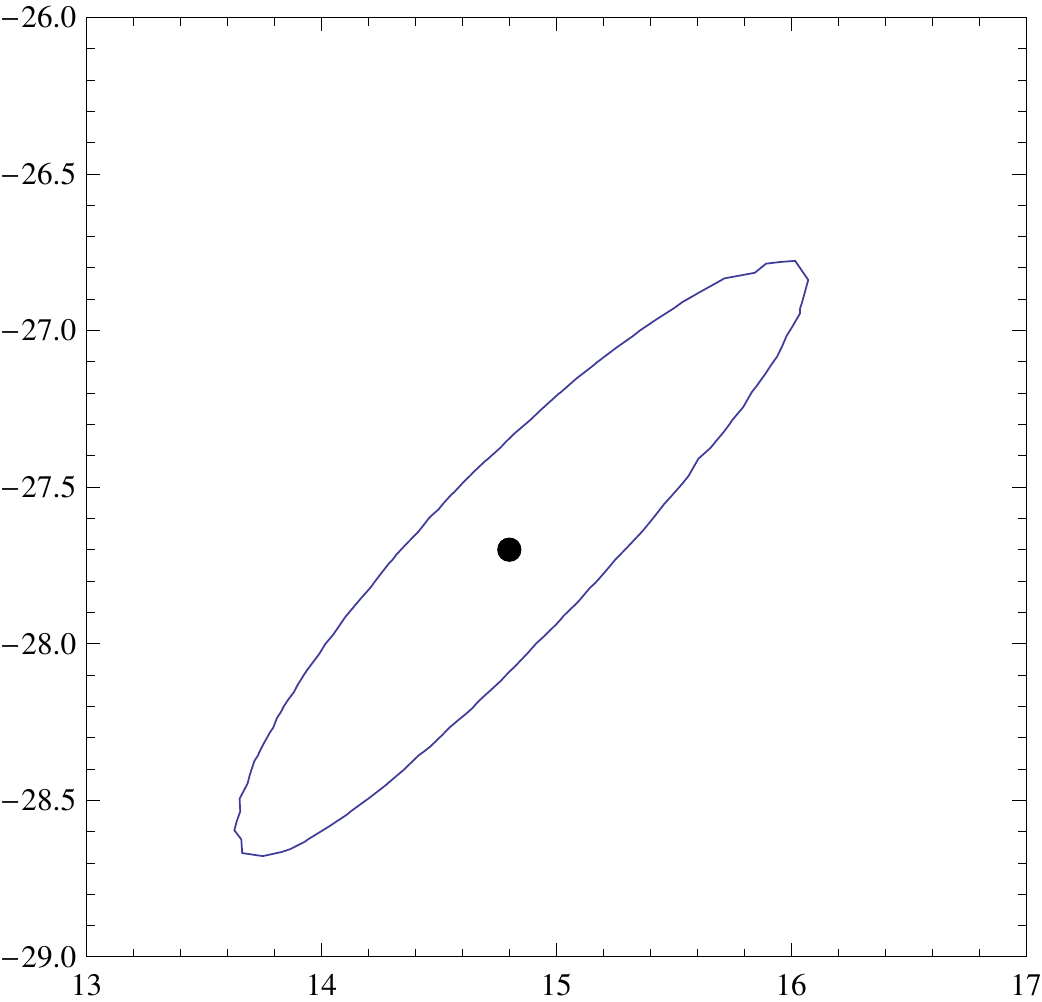}\\
\raisebox{1.5cm}{$G$}&\includegraphics[height=3cm,width=3cm]{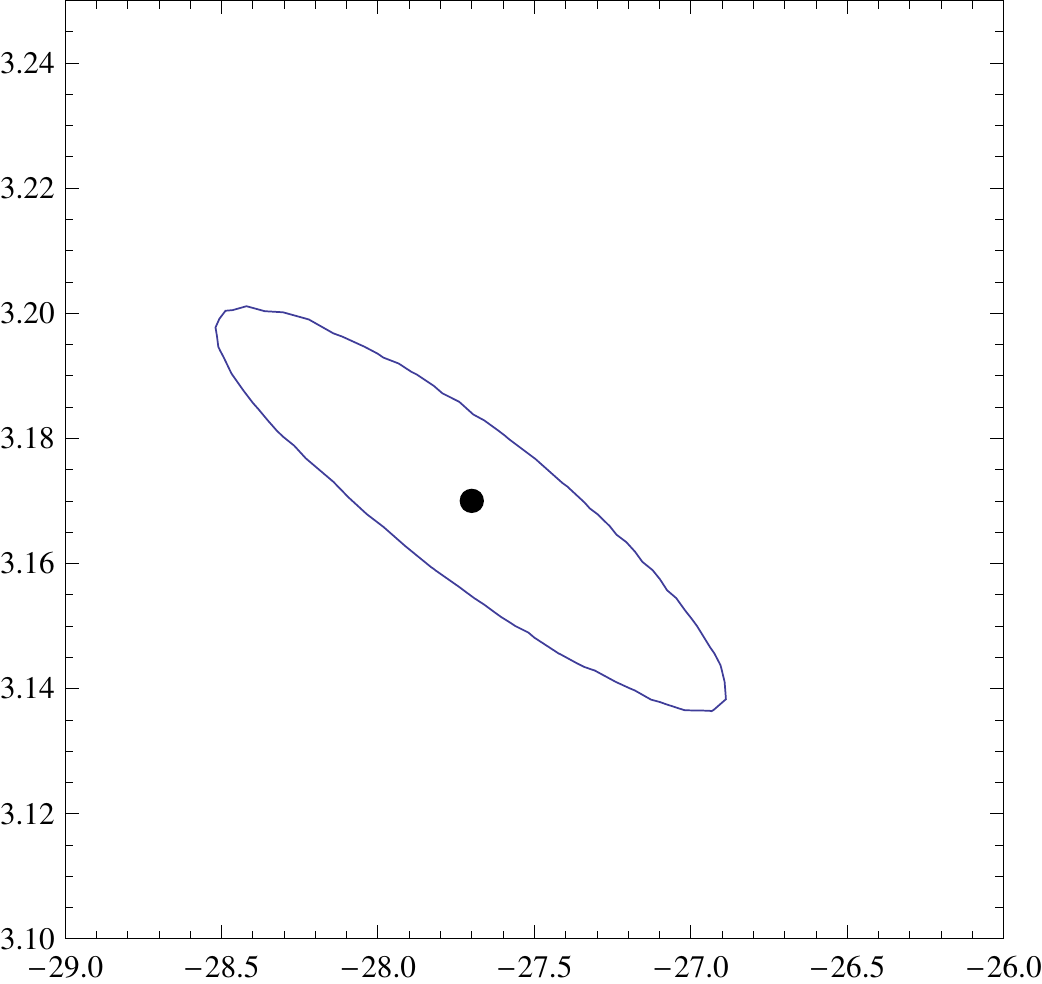}&\raisebox{1.5cm}{$\not$}&\includegraphics[height=3cm,width=3cm]{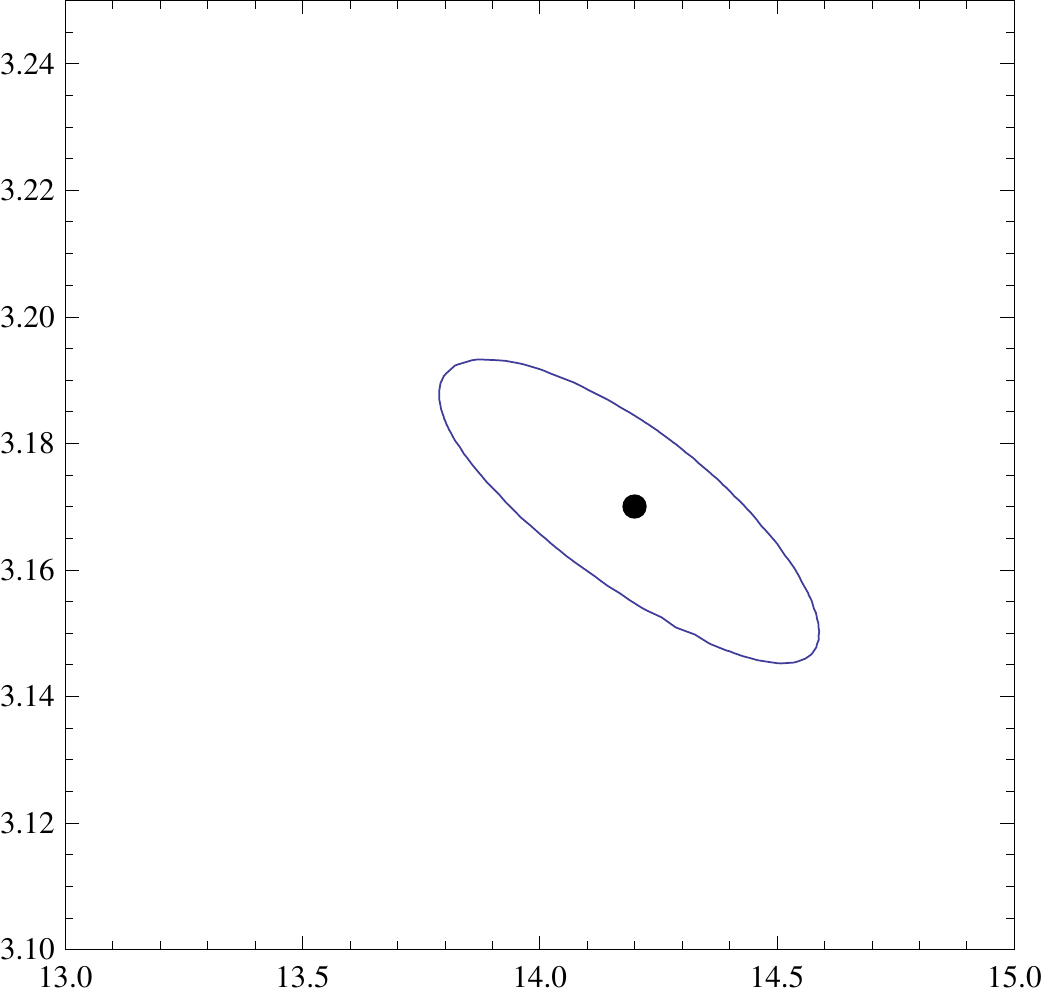}&\includegraphics[height=3cm,width=3cm]{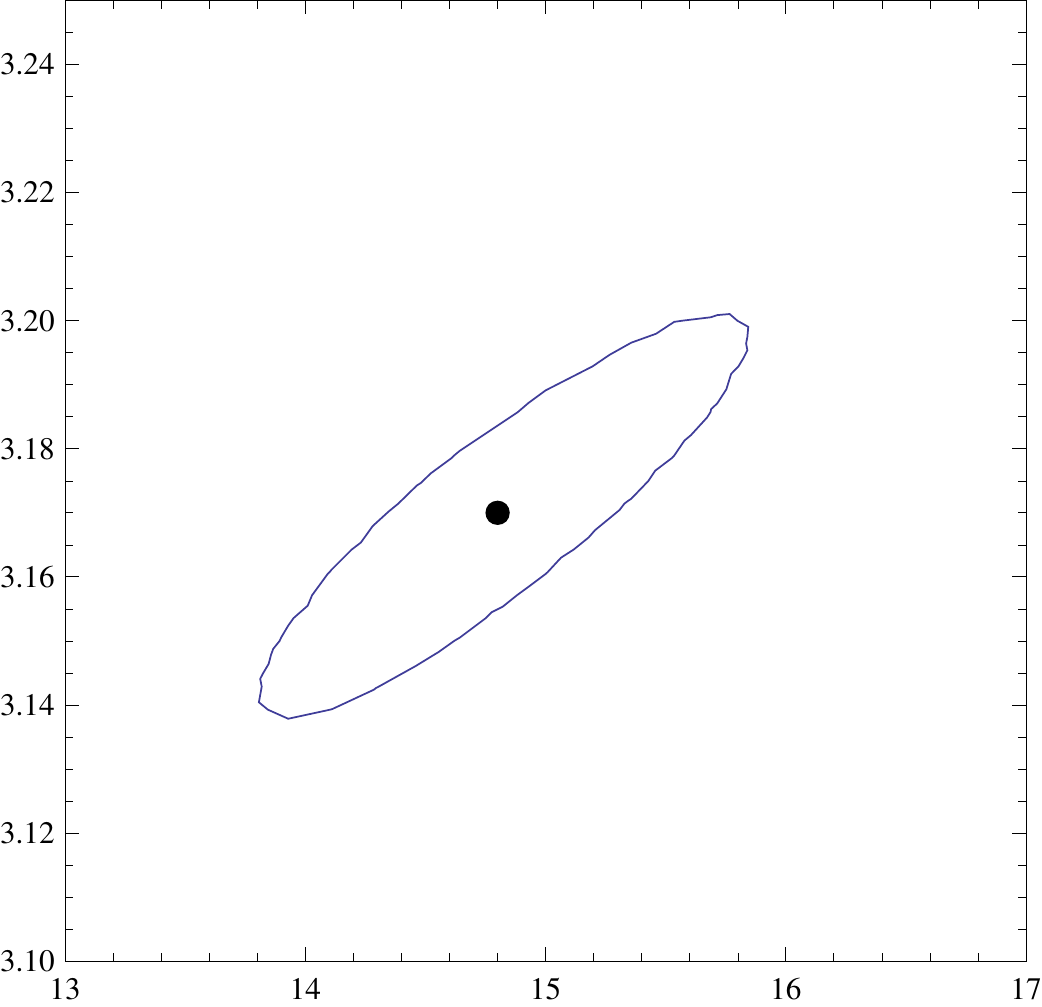}\\
\raisebox{1.5cm}{$\mu$}&\includegraphics[height=3cm,width=3cm]{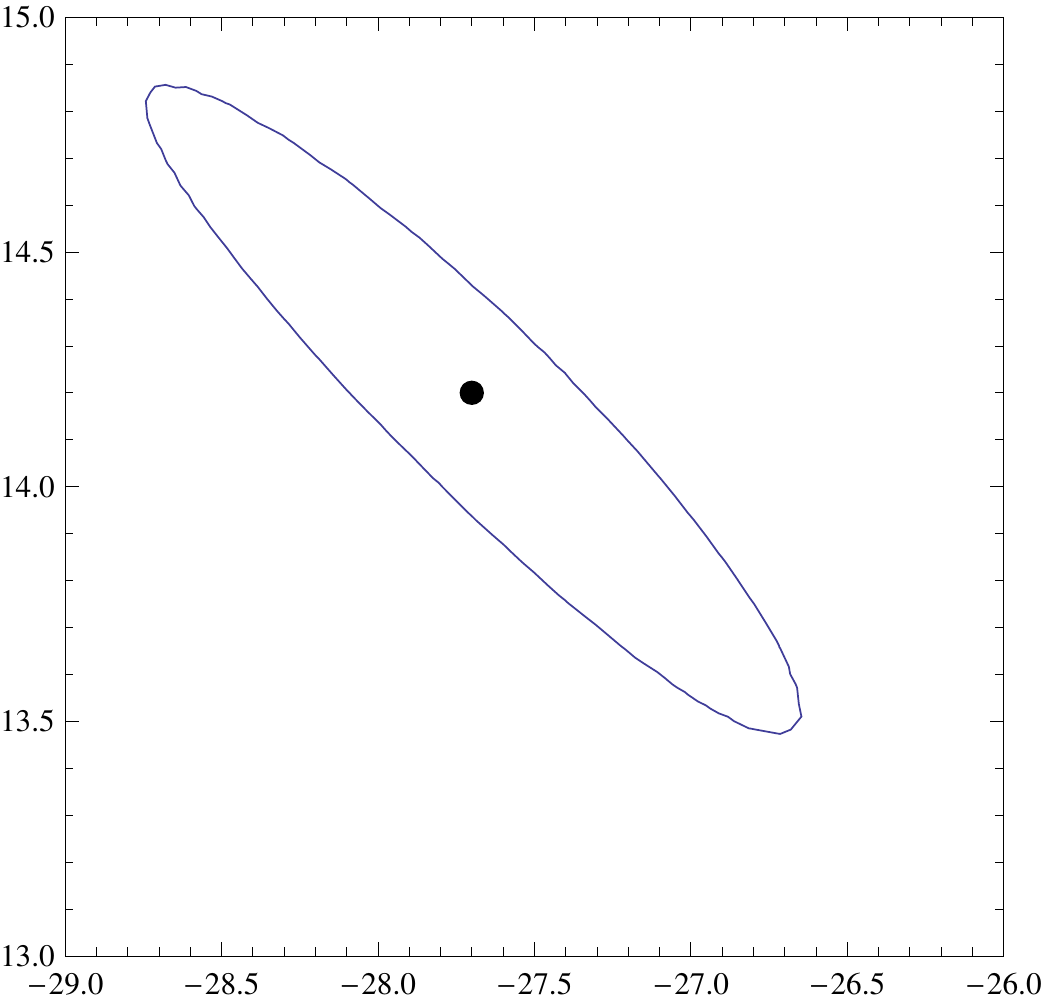}&\includegraphics[height=3cm,width=3cm]{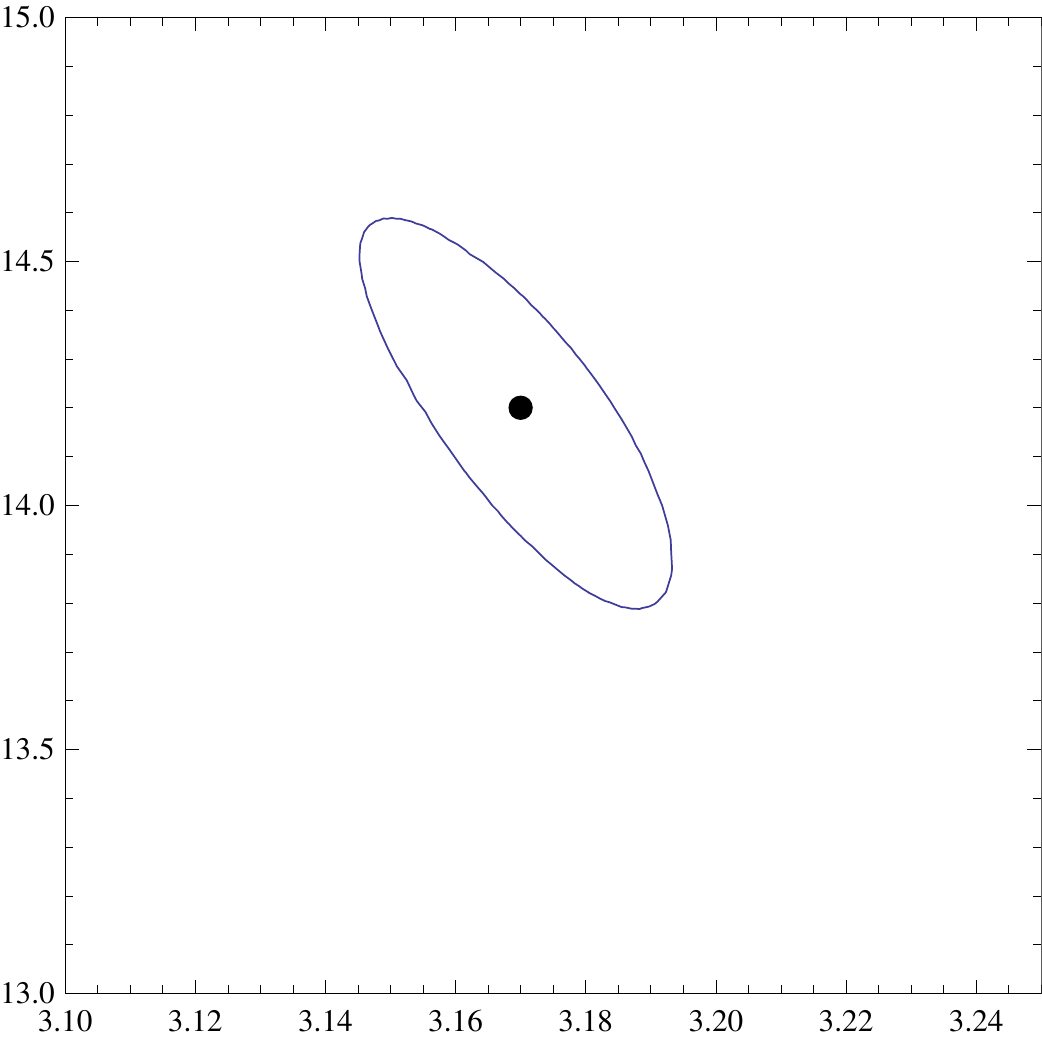}&\raisebox{1.5cm}{$\not$}&\includegraphics[height=3cm,width=3cm]{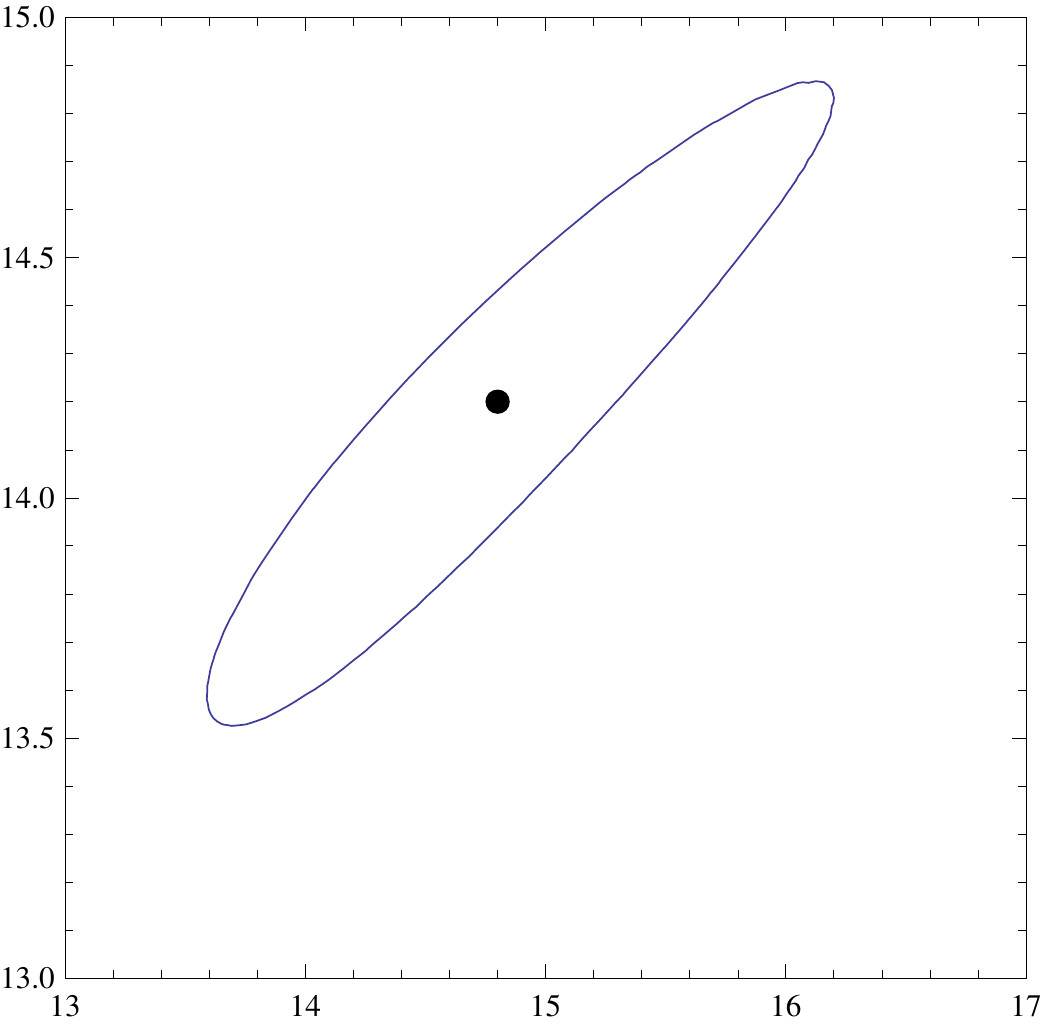}\\
\raisebox{1.5cm}{$F_6$}&\includegraphics[height=3cm,width=3cm]{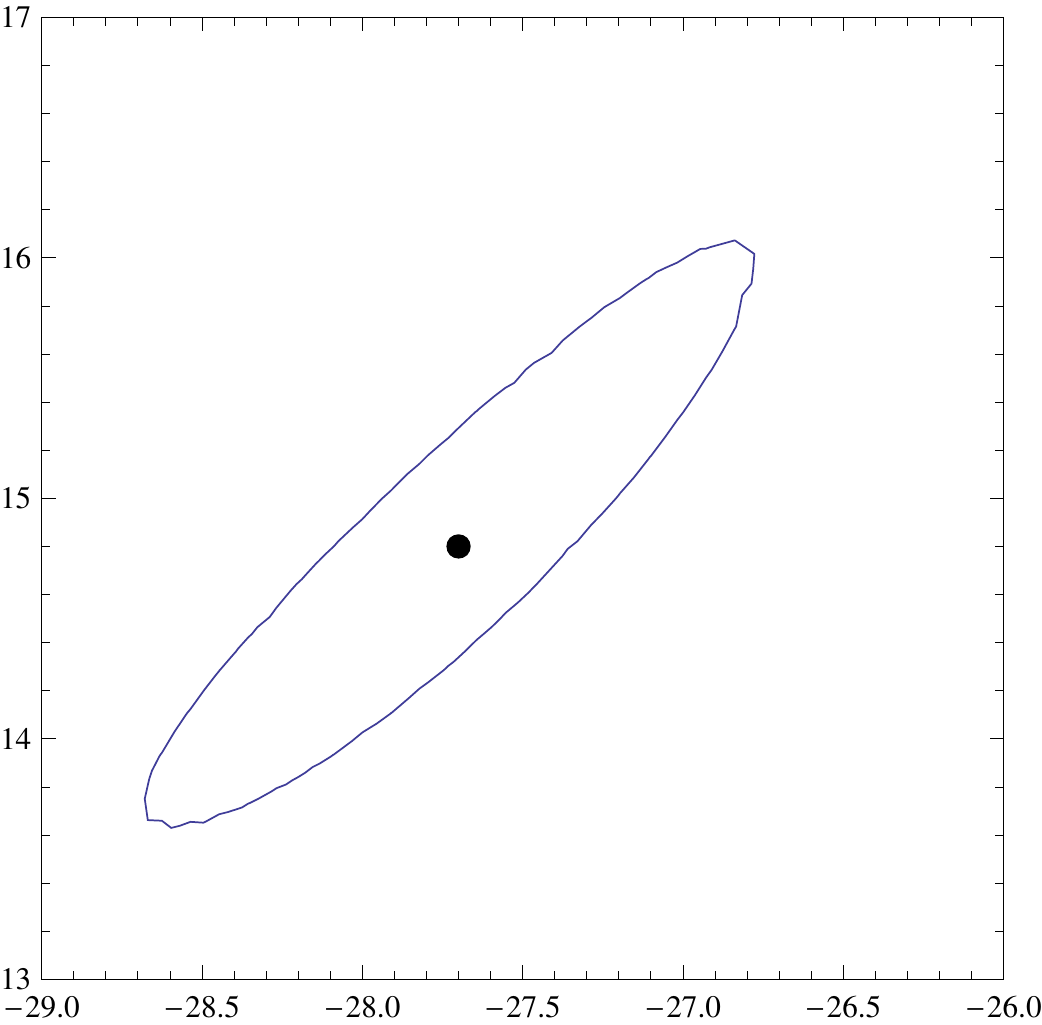}&\includegraphics[height=3cm,width=3cm]{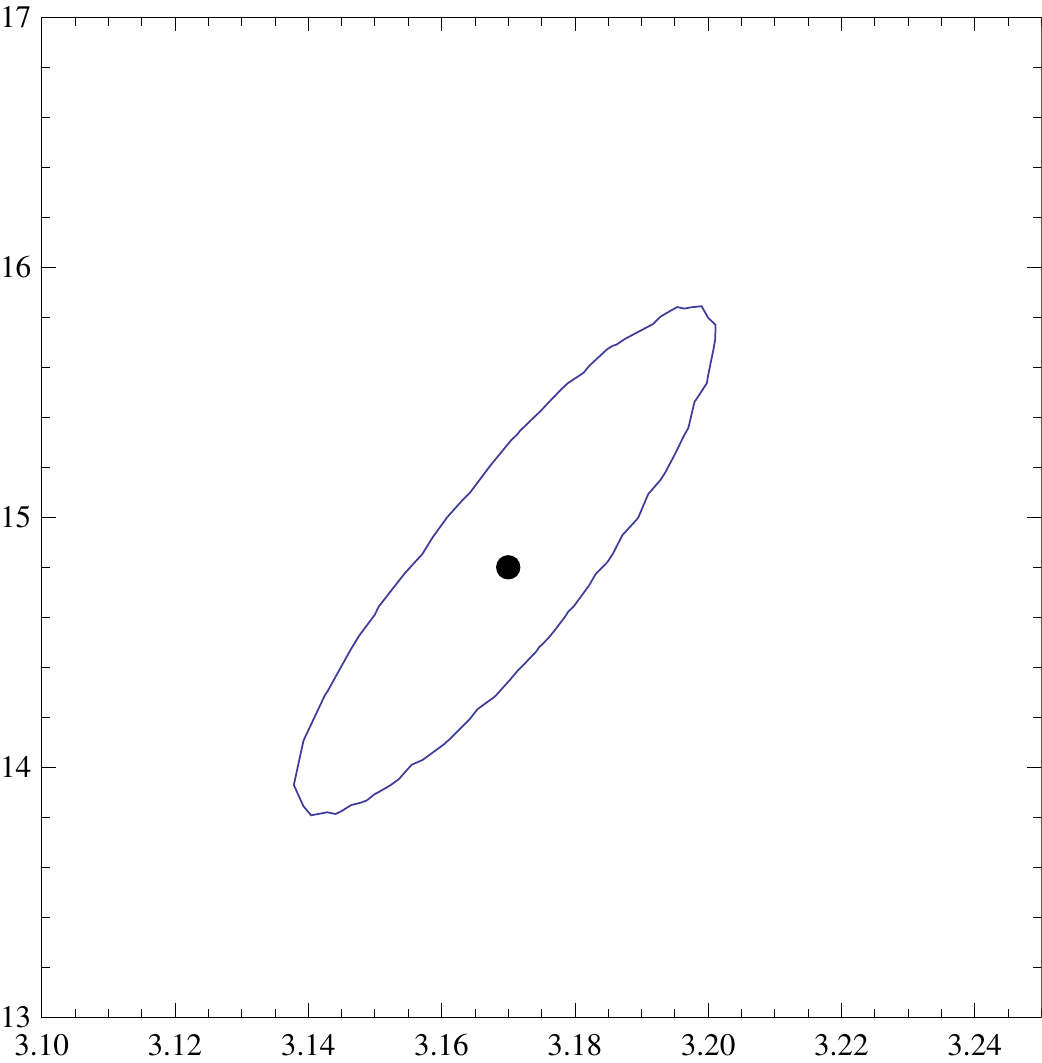}&\includegraphics[height=3cm,width=3cm]{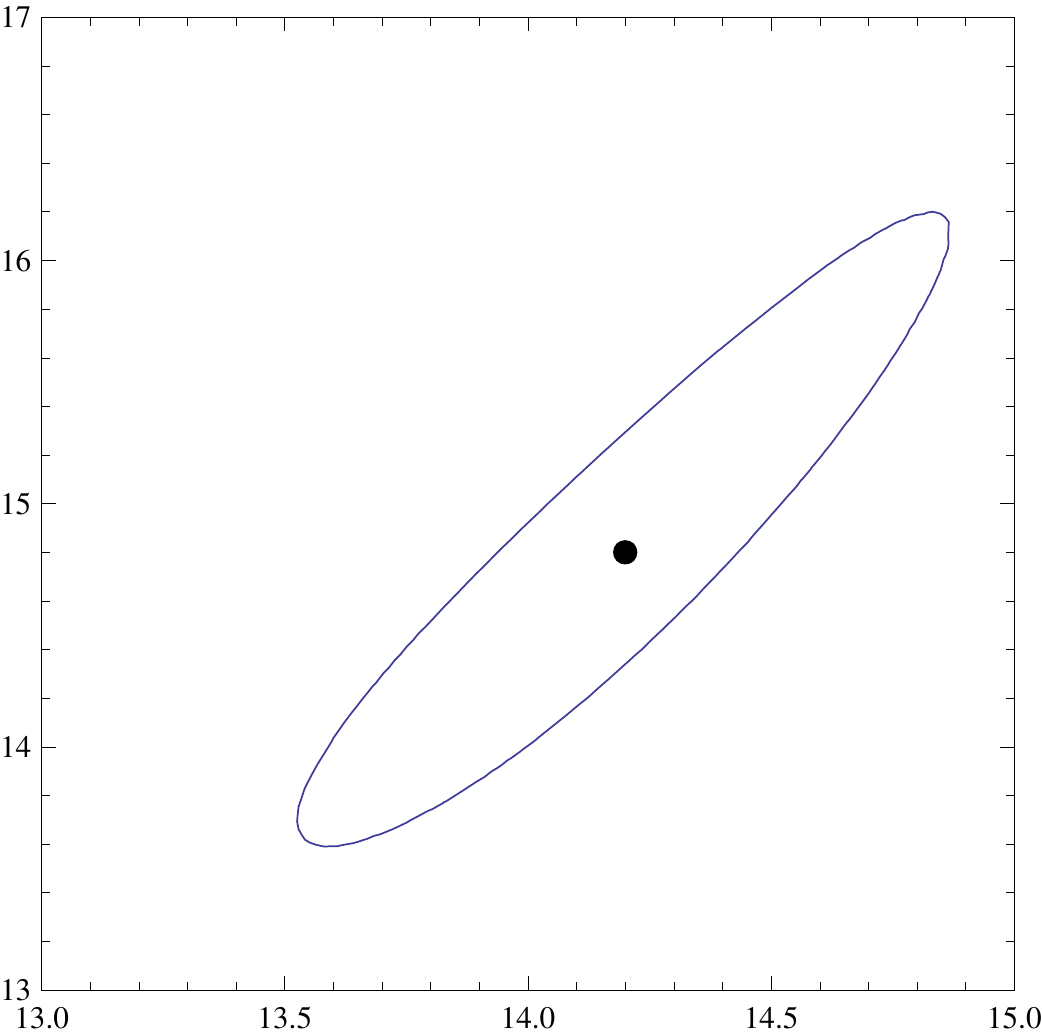}&\raisebox{1.5cm}{$\not$}
\end{array}$
\caption{Projections of the 6-dimensional 95\% confidence region into planes spanned by several pairs of parameters. Plots in each column share the same x-axis parameter as indicated at the top of each column. Plots in each row share the same y-axis parameter as indicated at the left of each row. Range for $F_1$ is (-29,-26); G: (3.1,3.25); $\mu$: (13,15); $F_6$: (13,17).}
\label{chisqu}
\end{figure}

The challenge is clearly that the low energy part and the $\Delta^+$ resonance range data points is difficult to fit well at the same time. The form factors (and thus the parameters $\mu$ and $G$) are generally functions of $k^2$ where $k$ is photon momentum. For real Compton scattering, $k^2$ is always 0, so the form factors should be constants in this paper. However in the case of the bare polarizabilities  it is possible that they vary with energy and/or scattering angle \cite{Schumacher2013}, which would make fitting with constant bare polarizabilities unsuccessful.

The strategy we propose to deal with the possible variation of bare polarizabilities is as follows. First, we fit only the peak range data points and fix the form factors(and $\mu$ and $G$) from this fitting. Then we fit the low energy data points varying only the bare polarizabilities. For the peak range, we use only the MAMI(2001) experiment \cite{MAMI2001}, which contains 436 data points with photon incident energy ranging from 260MeV to 455MeV. A good fit is achieved at $F_1=-27.7$, $G=3.17$, $\mu=14.2$, $F_6=14.8$, $\alpha_B=2.1$, $\beta_B=-8.1$ with $\chi^2\sim 830$. It is notable that $F_1$, $F_6$, $\mu$ and $G$ have not changed much from the complete fit of all data points, yet $\chi^2$ per datapoint is much smaller. This may be indicative of the fact that experimental data prior to this latest and more precise measurements may not be consistent with each other. In the past, one of the strategies for dealing with this has been to allow rescaling the crosssection data for each experiment within the systematical uncertainty which tends to be large \cite{Baranov:2000na}. 

The inclusion of the sigma channel and/or variation of the mass and width of the sigma meson do not appreciably alter the picture or the values of the best-fit parameters.

In Fig.\ref{chisqu}, we give the contour plots of $\chi^2$ with respect to several pairs of parameters for this fit. It is seen that $G$ is very strictly constrained.

\begin{figure}[htdp]
\setlength{\unitlength}{1cm}
\includegraphics[height=4cm,width=5cm]{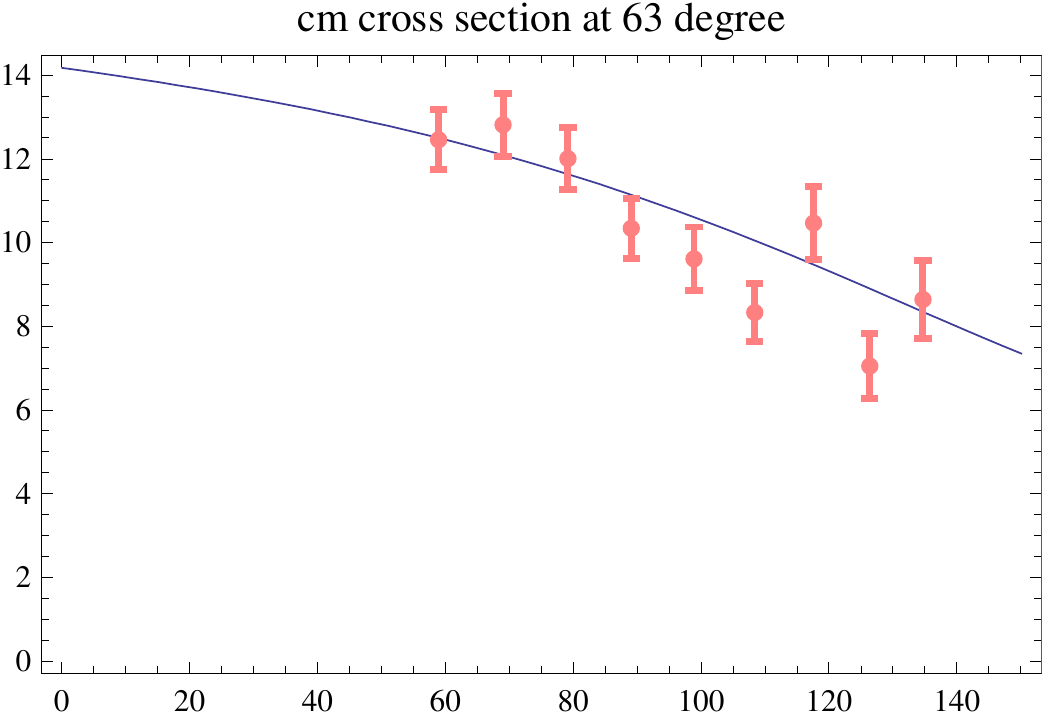}
\includegraphics[height=4cm,width=5cm]{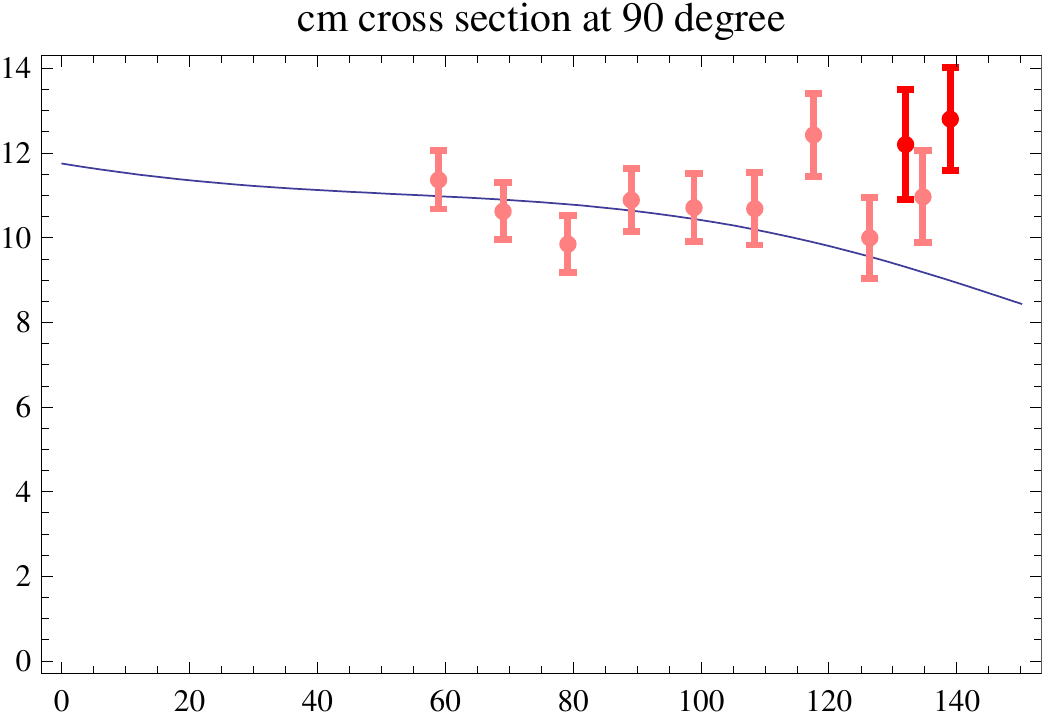}
\includegraphics[height=4cm,width=5cm]{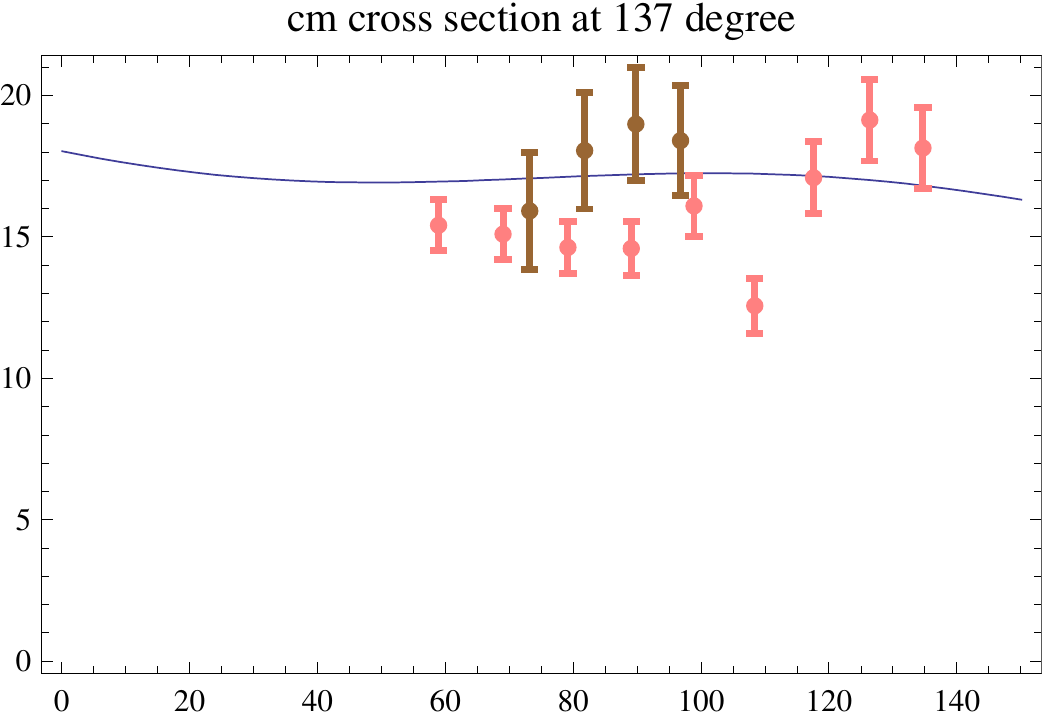}
\caption{Fixed c.m. angle cross section and the data points with photon incident energy below 140MeV, for the fit parameters of the low energy data points. The x-axis is lab frame photon energy and y-axis is c.m. frame differential cross section in units of nanobarn. The angles of the data points included may differ from the nominal by at most 2.5 degree. %Those data measured or recorded in lab frame have been converted to c.m. frame using the transition in eq.(\ref{labcmrela}). 
For \cite{Hallin1993,Baranov1974,Zieger92,Hunger97,Blanpied01,MAMI2001,Olmos2001,MacGibbon1995}, we use colors: Green, Blue, Black, Brown, Red, Gray, Pink and Yellow respectively.}
\label{lowangleplot}
\end{figure}

\begin{figure}[htdp]
\setlength{\unitlength}{1cm}
\includegraphics[height=4cm,width=5cm]{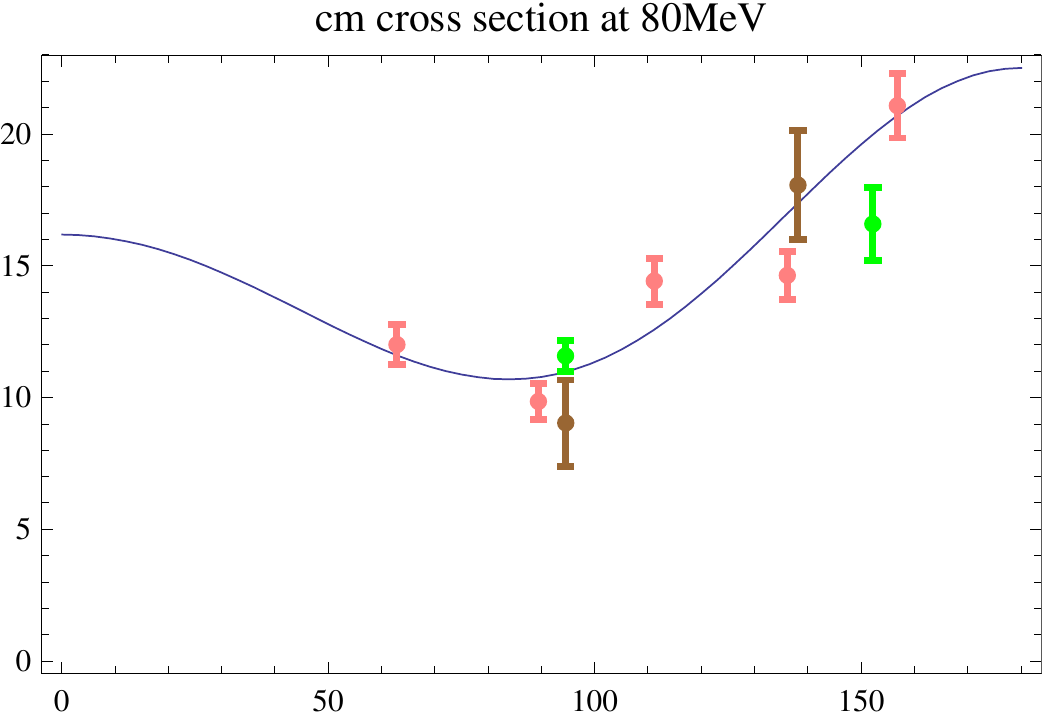}
\includegraphics[height=4cm,width=5cm]{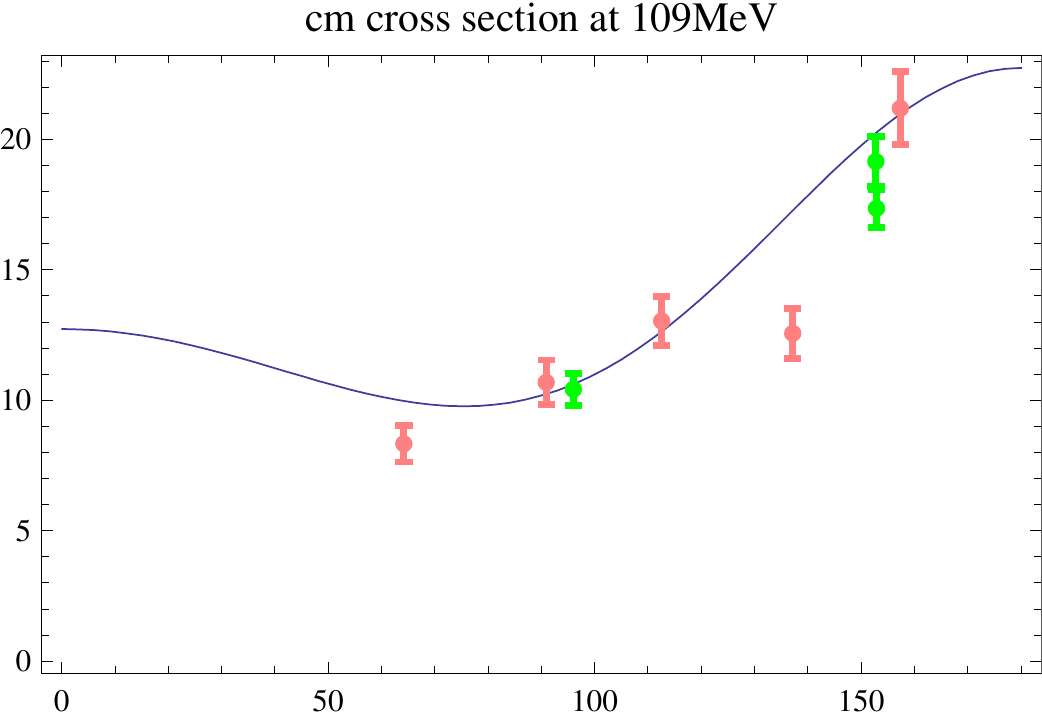}
\includegraphics[height=4cm,width=5cm]{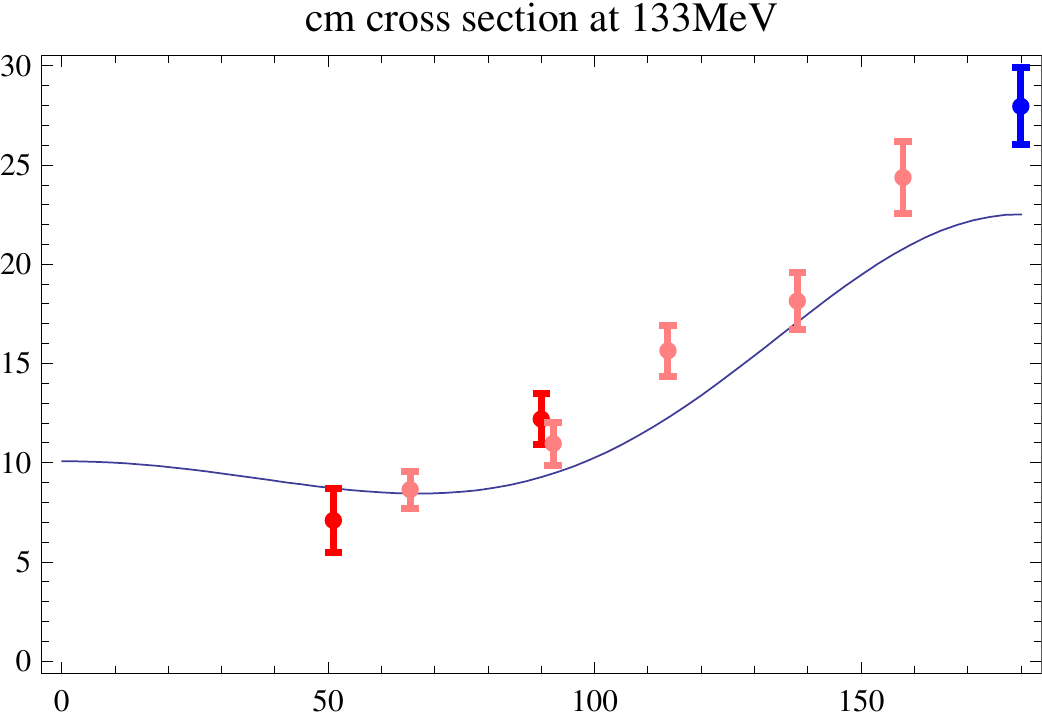}
\caption{Fit cross section and data points with photon incident energy below 140MeV, for the fit parameters of the low energy data points. The x-axis is c.m. frame scattering angle and y-axis is c.m. frame differential cross section in units of nanobarn. The incident energy of the data points included may differ from the values claimed by at most 2.5MeV. %Those data measured or recorded in lab frame have been converted to c.m. frame using the transition in eq.(\ref{labcmrela}). 
For \cite{Hallin1993,Baranov1974,Zieger92,Hunger97,Blanpied01,MAMI2001,Olmos2001,MacGibbon1995}, we use colors: Green, Blue, Black, Brown, Red, Gray, Pink and Yellow respectively.}
\label{lowenergyplot}
\end{figure}

We then use these values for $F_1$, $F_6$, $\mu$ and $G$ to fit the low energy data points, varying only $\alpha_B$ and  $\beta_B$. We take 68 data points with photon incident energy below 140MeV and obtain the best fit values of $\alpha_B=-4.6$ and $\beta_B=17.9$ with $\chi^2\sim 194$. See Fig. \ref{lowangleplot} and \ref{lowenergyplot}. At these values, $\bar\alpha+\bar\beta=11.3\pm 0.9\pm 2.3(95\% C.L.)$ and $\bar\alpha-\bar\beta=7.8\pm 3.3\pm 2.0(95\% C.L.)$, much closer to values extracted (from the same data) previously. The first error is determined from the MAMI fit, by investigating how the contribution of $F_1$, $G$ and $\mu$ to the polarizabilities varies in the 3-dimensional 95\% confidence region spanned by these three parameters. The second error is from the low energy fit.

\begin{table}[htdp]
\begin{center}
$\begin{array}{|c|c|c|c|c|c|c|}
\hline&F_1&G&\mu&F_6&\alpha_B &\beta_B\\ \hline
\mbox{whole fit}&-27.5&3.13&14.2&12.9&7.5&-8.2\\ \hline
1 \sigma &1.4&0.04&0.5&0.9&1.1&0.9\\ \hline
%90\%(\mbox{not } 2 \sigma) &2.9&0.09&1.1&1.6&2.3&1.5\\ \hline
95\% &2.8&0.08&1.1&1.7&2.2&1.6\\ \hline
\hline
\mbox{MAMI fit} &-27.7&3.17&14.2&14.8&2.1&-8.1 \\ \hline
1 \sigma &0.7&0.03&0.4&0.8&0.8&0.5\\ \hline
%90\% &1.1&0.04&0.6&1.3&1.2&0.9 \\ \hline
95\%&1.4&0.05&0.7&1.5&1.5&1.1 \\ \hline
\hline
\end{array}$
\end{center}
\begin{center}
$\begin{array}{|c|c|c|c|}
\hline
&\mbox{value}&\mbox{95\% error from MAMI fit}&\mbox{95\% error from low energy fit}\\ \hline
\bar\alpha+\bar\beta &11.3&0.9&2.3\\ \hline
\bar\alpha-\bar\beta & 7.8 &3.3&2.0\\ \hline
\end{array}$
\end{center}
\caption{Our best-fit parameter values and confidence regions}
\end{table}

\section{Discussion and Acknowledgements}

% One may well ask whether there is a description incorporating isospin. Indeed, the proton is part of an isospin I=1/2 nucleon multiplet , incorporating the proton and the neutron, while the Delta is part of the I=3/2 quadruplet containing $\Delta^{++}, \Delta^{+}, \Delta^{0}$ and $\Delta^{-}$. It is not  necessary to invent any new explanation for the fact that, for example, the system consisting of three u-quarks cannot exist in an s-wave state with spin 1/2 - the reason is that the total wavefunction must be antisymmetric under the exchange of the identical quarks and the color is always anti-symmetric for baryons.
%One possible way to incorporate this is that we may add a triple isospin index to the Rarita-Schwinger vector-spinor, $\psi^{\mu\alpha}_{abc}$ where $a,b,c=\pm1/2$, similar to what was done by Peccei. Then the restrictions due to antisymmetry can be imposed at the level of this larger state vector. Practical utility of such a description is limited by the unavailability of any data regarding the electromagnetic properties of the other $\Delta$ resonances, particularly the $\Delta^{++}$.
 
 Our model incorporates both minimal and non-minimal couplings. The former takes into account that the $\Delta^+$ is charged. The presence of the non-minimal couplings makes the description necessarily complicated, nevertheless the number of free parameters has been kept low. The parameters have clear physical interpretation, namely as the proton and $\Delta^+$ magnetic moment, as well as the strength of the $N\Delta$ magnetic transition (M1) and the two bare polarizabilities. We found that the data near the peak is well fit with our preferred set of parameters in the resonance region, but that the same set of parameters does not well describe the data at low energy. We have dealt with this in manner similar to \cite{McGovern:2012ew}.
 
Although we have been able to extract a value for the $\Delta^+$ magnetic moment, we cannot have high confidence in this value since proton Compton scattering does not probe the $\gamma\Delta^+\Delta^+$ vertex directly. The reason that this is possible at all is that the form factors have definite properties under Lorentz transformations, so that some linear combination of parameters which affects the Compton process also determines the magnetic moment of the $\Delta^+$. Qualitatively, we found that changes in the $\mu_{\Delta^+}$ affect the predicted cross-secion asymmetrically: lower values of the magnetic moment do not drastically change the prediction, but higher values greatly enhance the crosssection, both on and off the resonance. Therefore, our result conservatively stated is that  we exclude any values of  the $\Delta^+$ magnetic moment $\mu_{\Delta^+}$ larger than about $14.2$. This should be compared to that extracted by MAMI \cite{Kotulla:2002cg} at $2.7^{+1.0}_{-1.3} (stat) \pm1.5 (syst) \pm 3(theor)$, but note that the quoted error is dominated by theoretical model uncertainty.
Our upper bound is also consistent with naive quark model expectations and some model calculations \cite{Linde:1995gr, Kotulla:2003pm, Kotulla:2008zz, Kotulla:2002cg, Kotulla:2002tx}. The more reliable path towards determining the $\Delta^+$ magnetic moment would be to extend the model to include pion form-factors and thus cover the case of $N+\gamma \to N+\gamma+ \pi$ scattering which was also very well measured by some of the very same experimental groups as the Compton data considered in the present work.

The authors would like to thank G. Georgiou and J.D. Vergados for their valuable comments. 

A Project Funded by the Priority Academic Program Development of Jiangsu Higher Education Institutions (PAPD).

\appendix
\section{$\Delta^+\to p+\gamma$ Decay Width}
\label{sec:decayamp}
Aside from proton Compton scattering, another process can be readily accounted for in this unified $N-\Delta^+$ electromagnetic theory, that is the decay $\Delta^+\to N+\gamma$.  The Feynman rules for this diagram were given in Section \ref{sec:min}.
%, except the wave functions $u_4$ for $\Delta^+$. 
%In actual calculation, it is convenient to take the $\Delta^+$ in the rest frame and we only need $\Delta^+$ rest frame wave functions.\par

%\begin{figure}[htdp]
%\setlength{\unitlength}{1cm}
%\center{\includegraphics[height=4cm,width=6cm]{picture3.png}}
%\caption{$\Delta^+\to p+\gamma$ Feynman diagram.}
%\label{picture3}
%\end{figure}

We label the momentum and polarization of $\Delta^+$ as $k_1, \sigma_1$, the produced photon $k_2$,$\lambda_2$ and the proton $k_3$,$\sigma_3$, the matrix element is:

\beq
{\mathcal A}_{\sigma_1,\lambda_2,\sigma_3}=e \ \ {\bar u_2}(k_3,\sigma_3) \ {\tilde\Gamma}^\mu \ u_4(k_1,\sigma_1)\ \epsilon_\mu^*(k_2, \lambda_2).
\eeq
It is of interest to find the decay width $\Gamma_{3/2}$ and $\Gamma_{1/2}$ for the final state helicity $\frac{3}{2}$ and $\frac{1}{2}$ respectively. %, where the helicity means the sum of the proton and photon helicities. %, or equivalently the helicity of the initial $\Delta^+$. 
Evaluating this amplitude, we obtain after
%We calculated these helicity decay widths as:
%\bea
%&&\Gamma_{3/2}=\frac{\ |\vec {k}_2|}{32\pi^2 m^2} \frac{1}{4} \sum\limits_{\sigma_1=\{3/2,-3/2\}} \sum\limits_{\lambda_2,\sigma_3} \int|{\mathcal A}_{\sigma_1,\lambda_2,\sigma_3}|^2 {\mbox d}\Omega,\nonumber\\
%&&\Gamma_{1/2}=\frac{\ |\vec{k}_2|}{32\pi^2 m^2}\frac{1}{4} \sum\limits_{\sigma_1=\{1/2,-1/2\}} \sum\limits_{\lambda_2,\sigma_3} \int|{\mathcal A}_{\sigma_1,\lambda_2,\sigma_3}|^2 {\mbox d}\Omega.
%\label{decaywidth}
%\eea
%
substituting the values of m, M: %, the numerical values are:
\bea
\Gamma_{3/2}=&& 0.0047 F_1^2+ 0.056 F_2^2 + 0.001 F_4^2 + 0.001 F_5^2 + 0.032 F_1F_2  + 0.004F_1 F_5  \nonumber\\
&&+0.0139 F_2 F_5 + 0.032 F_1  + 0.1113 F_2 +  0.0139 F_5 + 0.0557,\nonumber\\
\Gamma_{1/2}=&&0.0004 F_1^2 + 0.0120 F_2^2 + 0.0002F_4^2 + 0.0002 F_5^2 + 0.0002 F_6^2 + 0.0058 F_1 F_2 \nonumber\\
&&+ 0.0005  F_1F_5  - 0.0006  F_1F_6  + 0.0037 F_2 F_5-  0.0039 F_2 F_6 - 0.0004 F_5 F_6\nonumber\\
&&+ 0.0043 F_1 + 0.0293 F_2  + 0.0027 F_5  - 0.0028 F_6 + 0.0108.
\eea

%Substituting the complex value of m in the amplitudes we get:
%\bea
%\Gamma_{3/2}=&& 0.0042 F_1^2+0.029 F_2 F_1-0.0001 F_4 F_1+0.0036 F_5 F_1+0.029 F_1+0.051 F_2^2\nonumber\\
%&&+0.0008 F_4^2+0.0008 F_5^2+0.10 F_2+0.013 F_2 F_5+0.013 F_5+0.051,\nonumber\\
%\Gamma_{1/2}=&&0.0004 F_1^2+0.0055 F_2 F1+0.00003 F_4 F_1+0.0005 F_5 F1-0.0004 F_6 F_1+0.004 F_1\nonumber\\
%&&+0.018 F_2^2+0.0002 F_4^2+0.0002 F_5^2+0.0001 F_6^2+0.027 F_2-0.0002 F_2 F_4\nonumber\\
%&&+0.0034 F_2 F_5+0.0025 F_5-0.0029 F_2 F_6+0.0001 F_4 F_6-0.0003 F_5 F_6-0.002 F_6+0.010.
%\eea

If, on the other hand, we let $m=M(1+x)$ then in the limit of small x we find
\bea
&&\Gamma_{3/2}\sim\alpha M x^3 |G|^2,\nonumber\\
&&\Gamma_{1/2}\sim\frac{\alpha M x^3 |G|^2}{3}.
\label{widthapprox}
\eea

%... On the other hand, this fit should account well for the $\Delta^+\to p+\gamma$ electromagnetic decay width. 
Experimentally $\Delta^+\to p+\gamma$ decay amplitudes can be extracted from the $\Delta^+$ peak of proton Compton scattering \cite{Blanpied01}
%\footnote{actually this is the only Compton scattering used to extract the $\Delta^+\to p+\gamma$ decay amplitudes as documented by PDG\cite{PDG}. Most often the amplitudes are extracted from $p+\gamma\to \pi+\gamma$. There are many more experiments for the latter process and the cross sections are large and precise.}. 
% Since our fitting is successful for the $\Delta^+$ resonance region, it is expected to predict $\Delta^+\to p+\gamma$ decay width. 
For the MAMI fit parameters  we can calculate $\Gamma_{3/2}=0.43MeV$ and $\Gamma_{1/2}=0.11MeV$, %described in Appendix \ref{sec:decayamp}, 
close to the values quoted in \cite{Beringer:1900zz,Blanpied01,Dugger:2007bt,Ahrens:2004pf,Arndt:2002xv}: $\Gamma_{3/2}=0.49-0.56MeV$ and $\Gamma_{1/2}=0.13-0.15MeV$.

\end{document}